\documentclass[1p]{elsarticle}

\usepackage{setspace}
\usepackage{xcolor}
\usepackage{graphicx}
\usepackage[labelfont=bf,labelsep=period,skip = 0pt]{caption}
\usepackage{xr-hyper}
\usepackage{chemformula} 
\usepackage[T1]{fontenc} 
\usepackage{multirow}
\usepackage{comment}
\usepackage{appendix}

\usepackage{xr}
\makeatletter

\newcommand*{\addFileDependency}[1]{
	\typeout{(#1)}
	\@addtofilelist{#1}
	\IfFileExists{#1}{}{\typeout{No file #1.}}
}
\makeatother

\journal{Fuel}

\begin{document}

\begin{frontmatter}

\title{Size distribution of primary submicron particles and larger aggregates in solvent induced asphaltene precipitation}

\author{Jia Meng}
\author{Somasekhara Goud Sontti}
\author{Mohsen Sadeghi}
\author{Gilmar F. Arends}
\author{Petr Nikrityuk}
\author{Xiaoli Tan\corref{cor1}}
\ead{xiaolit@ualberta.ca}
\author{Xuehua Zhang\corref{cor2}}
\ead{xuehua.zhang@ualberta.ca}
\address{Department of Chemical and Materials Engineering, University of Alberta, Alberta T6G 1H9, Canada}

\cortext[cor1]{Corresponding author}
\cortext[cor2]{Corresponding author}

\begin{abstract}
Asphaltene precipitation is a crucial phase separation phenomenon in the oil industry, especially in paraffinic froth treatment to extract bitumen from oil sands ores. This work reveals the formation of particles at 0.2 to 0.4 $\mu m$ in radius, defined as a primary sub-micron particle (PSMP), which is ubiquitous from diffusive mixing between asphaltene solution and any of 23 types of precipitants examined in our experiments. The yield and quantity of asphaltene particles are affected not only by the Hildebrand solubility parameter of the precipitants but also by the diffusion coefficients of the asphaltene solution and the precipitant. The Population Balance Model (PBM) with the Hildebrand solubility parameter has been used to model particle size distribution. Good agreement has been achieved between numerical predictions and the experimental data. It indicates that the colloid theory can describe the size distribution of PSMP and larger aggregates. Therefore, this study provides new insight into the mechanism for the dependence of yield and size distribution of the precipitated asphaltene particles on the composition of precipitants and adding inhibitors. Controlling of asphaltene yield and size distribution may be applied to the process of hydrocarbon separation or asphaltene precipitation prevention.

\end{abstract}


\end{frontmatter}


\section{Introduction}

Induced precipitation is a common process for the extraction and separation of certain compositions from a mixed source. Asphaltene precipitation is utilized in paraffinic froth treatment (PFT) units to remove water and solids impurities in the crude oils \cite{gray2015upgrading}. Asphaltene is enriched on polyaromatic cores \cite{chen2021ex} and defined as the components of crude oil that are insoluble in n-alkane solvents (i.e., n-pentane or n-heptane) and soluble in aromatic solvents (i.e., benzene or toluene) \cite{xu2018}.  On the other hand, asphaltene precipitation during the transportation of crude oil may cause severe problems such as blockage of pipelines \cite{Joshi2001,li2017experimental}. Temperature \cite{ARCINIEGAS2014202}, pressure \cite{Hirschberg1984}, gas injection \cite{ZANGANEH2018633,ZANGANEH2015132}, asphaltene concentration \cite{Haji2014}, solvent (precipitant) type \cite{VILASBOASFAVERO2017267,ENAYAT2020116250,AKBARZADEH2005159}, and solvent to crude oil ratio \cite{VILASBOASFAVERO2017267,AKBARZADEH2005159,ROGEL2017271,Maqbool2011,Ramos2020,duran2019} have been shown to play significant roles to control the kinetic of asphaltene precipitation. 

In general, the yield of asphaltene increases with the solvency effect of precipitant, which can be achieved by increasing the concentration of alkanes and changing to shorter chain-length alkanes \cite{ROGEL2017271}. From a solution viewpoint, the increase of the solid-liquid equilibrium ratio from solvency effect results in asphaltene changes from a liquid phase to a solid phase \cite{RODRIGUEZ2019116079}. From a colloidal viewpoint, asphaltene colloids aggregates disperse in crude oils and are stabilized by the steric repulsion of the extended structure \cite{gray2011,wang2010,wang2009}. Changing surrounding conditions, such as medium composition \cite{maqbool2009}, induces the collapse of the extended structure. The steric repulsion between asphaltene colloids reverses to van der Waals attraction \cite{wang2010,wang2009}, under which asphaltene colloids aggregate to grow larger and manifest as phase separation when solvent to bitumen (S/B) ratio is higher than onset. The collision efficiency between asphaltene particles increases with the solvency effect \cite{wang2010}. Small asphaltene particles grow large and appear as a new phase. Either viewpoint indicates high solvency effect is conducive to the destabilization and aggregation of asphaltene. Besides, asphaltene precipitates are porous and tenuous fractal structures. Settling behavior suggests that the morphology of asphaltene precipitate is also affected by the precipitant composition. The settling rate of asphaltene particles in pentane is two magnitudes higher than in heptane \cite{Casas2019}, indicating a higher fractal dimension and larger aggregates.

Adding inhibitors can inhibit asphaltene precipitation in general \cite{balestrin2019}. However, for some specific cases, adding inhibitors may have the opposite function that to enhance asphaltene precipitation \cite{HUFFMAN2021121320}. The importance of inhibitors on destabilization and aggregation of asphaltene particles needs to be further investigated. In addition to precipitant composition (thermodynamic aspect), recent works show the growth dynamics of an individual domain during dilution-induced phase separation is also determined by the temporal and spatial characteristics of the mixing (i.e., hydrodynamics aspect) \cite{meng2020,Zhang2015}. This is because mixing directly affects local concentration, which can significantly affect the early stage of phase separation \cite{meng2020,Zhang2015}. 


Despite the importance of the precipitant on asphaltene precipitation, most studies have used a batch system to mix bitumen or other model oils with precipitant from which an aliquot is picked to observe \cite{xu2018,li2017experimental,maqbool2009}. Although these setups can provide information on the morphology of the precipitated asphaltene, it is difficult to decouple the effects from solubility parameters, diffusion, or external mixing.


Recently, micro platform device has been leveraged to study asphaltene precipitation \cite{sieben2015,sieben2016,mozaffari2021}. We developed a quasi-2D microfluidic chamber to study asphaltene precipitation in-situ \cite{meng2021primary,meng2021microfluidic}. Unlike using a mixer to control mixing dynamics in a bulk system, the mixing in the microchamber was controlled by diffusion in the confined space inside the quasi-2D channel \cite{lu2017universal}. The precipitated asphaltene can be in-situ visualized by a total internal reflection fluorescence microscope (TIRF) with a high spatial resolution ($\sim$ 200 $nm$) through an opaque medium containing asphaltene \cite{Dyett2018Growth,Dyett2018Coalescence,Dyett2020}. We found the presence of primary sub-micron particles (PSMP) with an equivalent radius of 0.2 to 0.4 $\mu m$ from asphaltene precipitation in pentane\cite{meng2021primary}. However, the ubiquitous formation and distribution of PSMP are still unknown in other types of precipitants and a real bitumen system.


The aim of this work is to understand size distribution of primary submicron particles and their aggregates in asphaltene precipitation induced by diffusive mixing with different precipitants. In total, 23 types of precipitants were examined, including three types of solvents pentol (pentane-toluene mixture), heptol (heptane-toluene mixture), and dectol (decane-toluene mixture), the mixture of heptane and decane, and the solutions containing an inhibitor. Population balance model (PBM) was established to relate the particle size distribution to the properties of the precipitant. The good agreement between the prediction from PBM model and the experimental results suggests that the colloid theory may describe the precipitation from PSMP to larger aggregates. It is worth noting that a real bitumen system was also investigated to compare the results from the model asphaltene in toluene solution. The findings of this work provide a further understanding of the solvent composition influence on asphaltene precipitation. The novelty of this work is development and validation of PBM model referring to particles size distribution for aggregates in solvent induced asphaltene precipitation.


\section{Experimental methods}

\subsection{Chemical and sample preparation}
Asphaltene was prepared from Murphy Oil (USA) pentane asphaltene (i.e., C5-asphaltene). Bitumen sample is an Athabasca bitumen supplied by Sycrude, Canada, Ltd. The elementary composition of the asphaltene sample can be found in our previous research \cite{meng2021primary}. Toluene (Fisher Scientific, ACS grade, 99.9\%+) was used as the solvent of asphaltene. n-Pentane (Fisher Scientific, 98\%), n-heptane (Fisher Chemical, 99\%), and n-decane (Fisher Scientific, 99.3\%+) were used as solvents. Nonylphenol (Aldrich, technical grade) was used as an inhibitor.

Asphaltene was treated following the same method reported in the previous study \cite{meng2021primary} to remove any inorganic solids. The treated asphaltene was dissolved in toluene to the concentration of 17 $g/L$, labeled as solution A. Bitumen sample was diluted by toluene to 100 $g/L$ and filtrated by the same method as asphaltene sample to remove any inorganic solids. The treated bitumen solution was used as solution A. With $\sim$ 17 $wt.\%$ C5-asphaltene in Athabasca bitumen \cite{Feng2013}, the asphaltene concentration in 100 $g/L$ bitumen solution was same as that in the toluene solution.

\subsection{Compositions and diffusion coefficients of solution B}
In total, 23 types of precipitants were used to investigate the solvency effect on asphaltene precipitation. 3 pure solvents were n-pentane, n-heptane, and n-decane. The paraffinic solvent was blended with toluene to prepare 9 types of solution B at the initial concentration ($\phi_0$) listed in Table \ref{table1}. 3 mixtures of heptane and decane with different concentrations were studied, as shown in Table \ref{table2}. Solution B as heptane was chosen to study asphaltene precipitation in a real bitumen system. The inhibitor nonylphenol at 4 concentrations of 10 to 10000 $ppm$ was added to both solution A (asphaltene in toluene solution) and solution B of pentane or heptane. The experiments followed the same procedure as above without nonylphenol in solution A or B.  

\begin{figure}[ht]
\centering
\includegraphics[width=1\columnwidth]{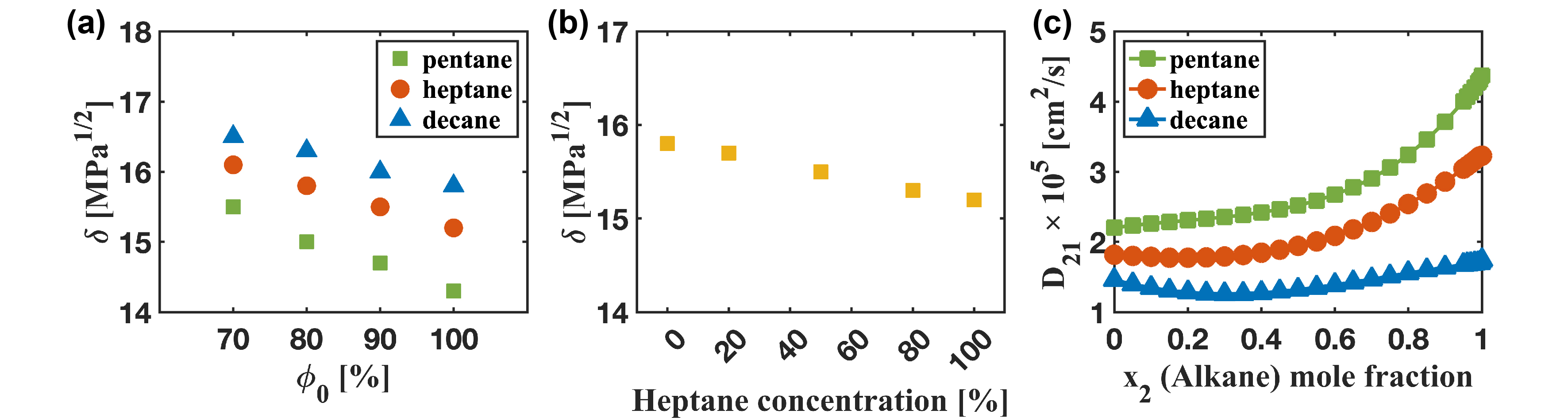}
\captionsetup{font={normal}}

\caption{Hildebrand solubility parameter and diffusion coefficient versus composition of the precipitant: a) mixture of alkanes and toluene, b) mixture of heptane, and decane. c) Diffusion coefficient of pentane, heptane, and decane in mixtures with
toluene.}
\label{HSP}
\end{figure}

Hildebrand solubility parameter ($\delta$) was used to quantify the solvency effect. For example, the larger of $\delta$ difference between asphaltene and solvent, the more asphaltene or the earlier of onset of asphaltene precipitates. $\delta$ for the pure groups are adopted from the literature \cite{barton2017crc}. $\delta$ for the mixtures are calculated based on the volume fraction of the components ($\phi_i$) and Hildebrand solubility parameter of each components ($\delta_i$), as shown in Equation (\ref{sol}) \cite{ANGLE2006492,wang2003asphaltene}. The relationship between $\delta$ and the composition of the precipitants is shown in Figure \ref{HSP}(a)(b). 
 
\begin{equation}
\label{sol}
\delta_{mixture}=\Sigma \delta_i \phi_i
\end{equation}

\begin{table}[ht]
\centering
\captionsetup{font={normal}}

\caption{Composition of the mixtures of n-alkanes and toluene.}
\label{table1}
\begin{tabular}{|c|l|c|l|c|l|}
\hline
\multicolumn{2}{|c|}{n-Pentane}                                                     & \multicolumn{2}{c|}{n-Heptane}                                                      & \multicolumn{2}{c|}{n-Decane}                                                       \\ \hline
$\phi_0^{pen}$ & \begin{tabular}[c]{@{}l@{}} $\delta$ ($MPa^{1/2}$) \end{tabular} & $\phi_0^{hep}$ & \begin{tabular}[c]{@{}l@{}} $\delta$ ($MPa^{1/2}$) \end{tabular} & $\phi_0^{dec}$ & \begin{tabular}[c]{@{}l@{}}$\delta$ ($MPa^{1/2}$)\end{tabular} \\ \hline
70\%  & 15.5                                                                        & 70\%  & 16.1                                                                        & 70\%  & 16.5                                                                        \\ \hline
80\%  & 15.0                                                                        & 80\%  & 15.8                                                                        & 80\%  & 16.3                                                                        \\ \hline
90\%  & 14.7                                                                        & 90\%  & 15.5                                                                        & 90\%  & 16.0                                                                        \\ \hline
100\% & 14.3                                                                        & 100\% & 15.2                                                                        & 100\% & 15.8                                                                        \\ \hline
\end{tabular}
\end{table}

\begin{table}[ht]
\centering
\captionsetup{font={normal}}

\caption{Composition of the mixtures of n-heptane and n-decane.}
\label{table2}
\begin{tabular}{|c|c|c|c|}
\hline
\multirow{3}{*}{} & \multicolumn{2}{c|}{Composition}                                         & \multirow{3}{*}{\begin{tabular}[c]{@{}c@{}}$\delta$ ($MPa^{1/2}$)\end{tabular}} \\ \cline{2-3}
                  & \multirow{2}{*}{n-Heptane (vol.\%)} & \multirow{2}{*}{n-Decane (vol.\%)} &                                                                                              \\
                  &                                     &                                    &                                                                                              \\ \hline
n-Heptane         & 100                               & 0                                  & 15.2                                                                                         \\ \hline
Mixture 1         & 80                                  & 20                                 & 15.3                                                                                         \\ \hline
Mixture 2         & 50                                  & 50                                 & 15.5                                                                                         \\ \hline
Mixture 3         & 20                                  & 80                                 & 15.7                                                                                         \\ \hline
n-Decane          & 0                                   & 100                                & 15.8                                                                                         \\ \hline
\end{tabular}
\end{table}

We compare the diffusion coefficients of alkanes in toluene as shown in Figure \ref{HSP}(c). The diffusion coefficient of pentane, heptane, and decane are simulated by UNIFAC (see details in Supporting information). Diffusion coefficient of pentane to toluene ($D_{pt}$) is higher than heptane ($D_{ht}$) and decane ($D_{dt}$) (i.e., $D_{pt}$ > $D_{ht}$ > $D_{dt}$). Mixing of alkane and toluene is a mutual diffusion process, affected by the concentration of toluene. Diffusion coefficient of the paraffinic solvent decreases during the mixing process. However, $D_{pt}$ > $D_{ht}$ > $D_{dt}$ holds for all of toluene concentration, in the particular at higher concentration of toluene.

\subsection{Detection and analysis of asphaltene precipitates}

\begin{figure}
\centering
\includegraphics[width=1\columnwidth]{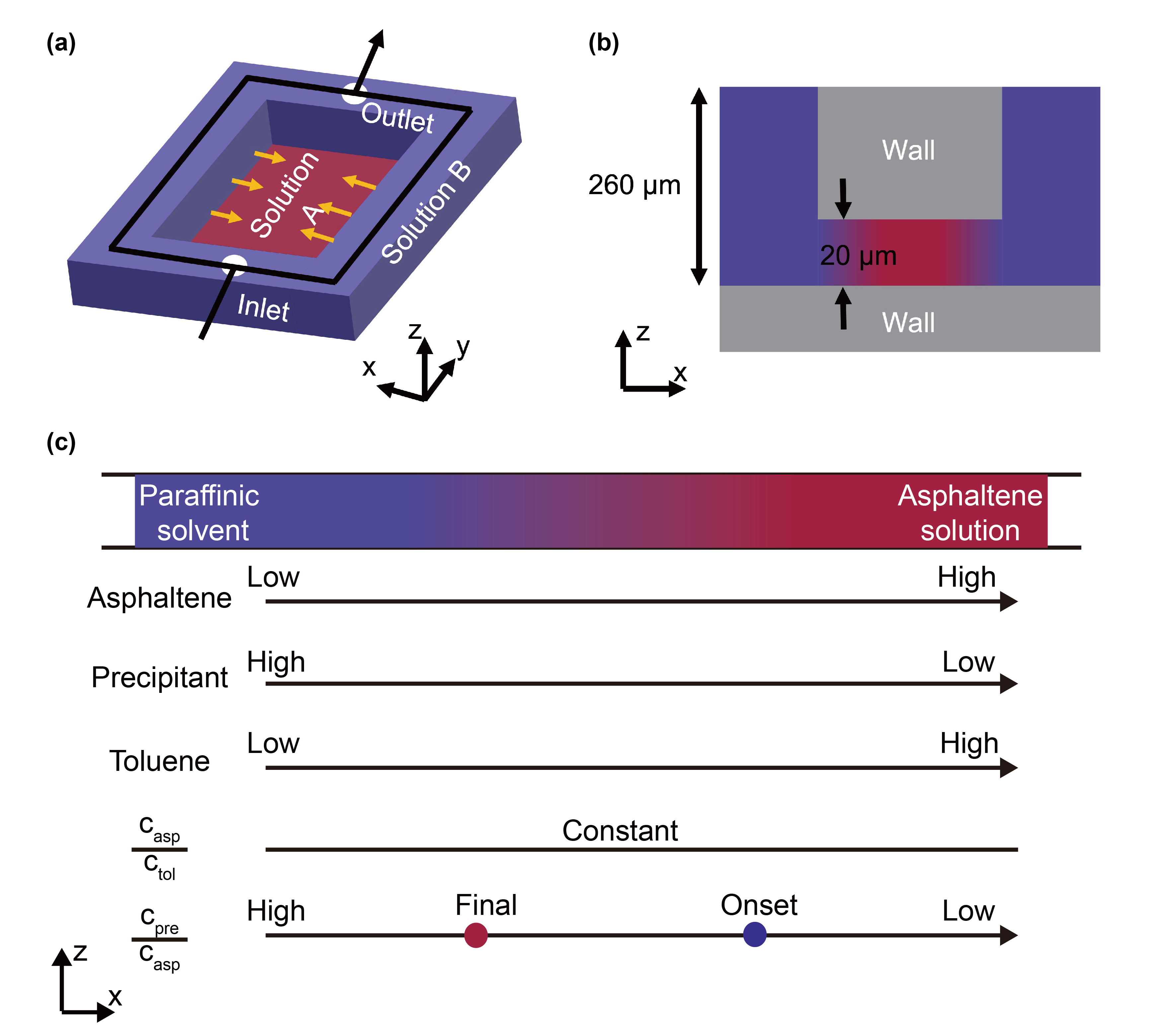}
\captionsetup{font={normal}}

\caption{a) Sketch of the microchamber used in this study. The black arrow indicates the flow direction of the precipitant. Yellow arrows indicate the diffusion direction of the precipitant. b) Side view of the cross-section of the microchamber. c) Schematic of the concentration of chemical composition in the diffusive mixing zone in the quasi-2D channel. Purple and red dots indicate the chemical composition
of onset and stop point for asphaltene precipitation, respectively.
( $\frac{c_{asp}}
{c_{tol}}$ ) is the ratio between asphaltene to toluene. ($\frac{c_{pre}}{c_{asp}}$) is the ratio between precipitant to asphaltene.}
\label{chamber}
\end{figure}

A quasi-2D microfluidic chamber was used to induce asphaltene precipitation by diffusive mixing with solution B, as shown in Figure \ref{chamber}(a)(b). More details of the quasi-2D microfluidic were described in our previous works \cite{meng2021primary,meng2021microfluidic}. In brief, solution A pre-filled the microfluidic chamber and solution B (the precipitant) was then injected through the deep side channels. Solution B transversely diffused into the main quasi-2D channel. Asphaltene precipitation began when the precipitant concentration reached the onset and finished when the concentration of asphaltene in the mixture was too low for precipitation, as sketched in Figure \ref{chamber}(c). The diffusion length was defined as 82.5 $\mu m$ in the direction of the concentration gradient (x-direction in Figure \ref{chamber}(a)).

We noted that the density difference between solution A and B did not influence the mixing dynamics in our confined quasi-2D channel, in contrast to a mixing process in a large bulk system. The buoyancy and gravity effects in our chamber can be estimated by a dimensionless number, Rayleigh number (Ra) \cite{Zhang2015}:
\begin{equation}
\label{Ra}
Ra=\frac{\Delta \rho g (h/2)^3}{\mu D_{21}}
\end{equation}
where $\Delta \rho$ is the density difference between alkanes and toluene, $g$ is the gravity acceleration constant, $h$ is the height of the chamber, $\mu$ is the dynamic viscosity of toluene, and $D_{21}$ is the diffusion coefficient of alkanes and toluene. The contribution of asphaltene on density and viscosity is negligible due to the small amount. Based on Equation (\ref{Ra}), Ra $\approx$ 2 for pentane and toluene (a combination with the largest density difference), which is three magnitudes smaller than the critical Ra (i.e., $Ra$ = 1708) \cite{Zhang2015}. Therefore, the mixing dynamics in all of our experiments were not affected by gravity.

Total internal reflection fluorescence microscope (DeltaVision OMX Super-resolution microscope, GE Healthcare UK Limited, UK) was used to detect the asphaltene particles. The images of asphaltene particles were captured after injecting solution B for five minutes. MATLAB (The MathWorks, Inc., US) was used for image analysis to get the particle size, surface coverage, and particle quantity. The details of the process for image analysis can be found in our previous work \cite{meng2021primary,meng2021microfluidic}. Briefly, surface coverage was normalized by the unit area in the field of view. Fractal aggregates were treated as single units rather than counting primary sub-micron particles when particle quantity was counted.

\section{Population balance model}
\begin{figure}[ht]
\centering
\includegraphics[width=1\columnwidth]{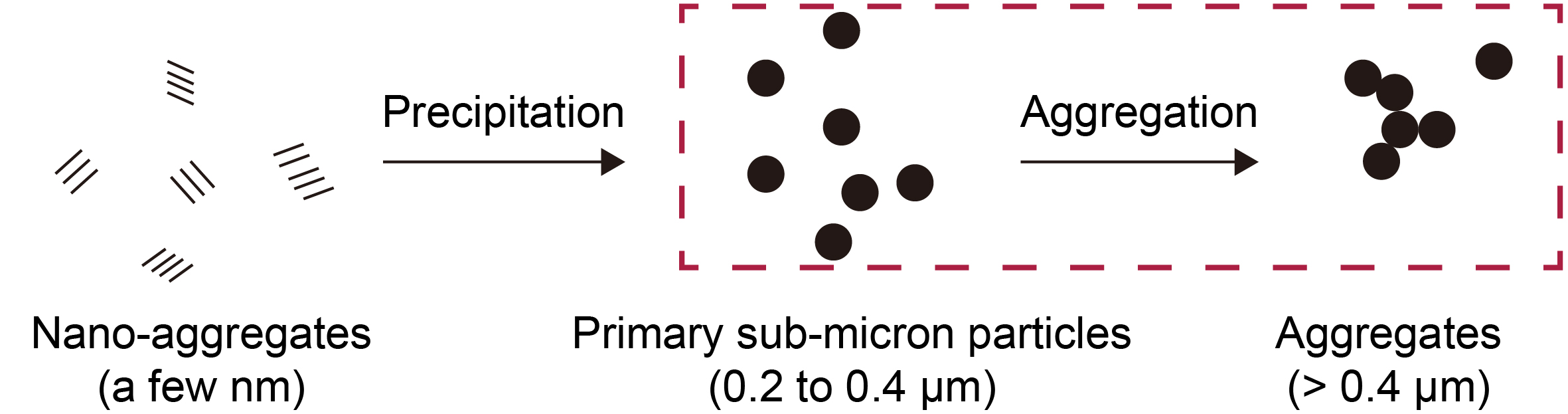}
\captionsetup{font={normal}}

\caption{Sketch of the formation of PSMP by asphaltene precipitation and the aggregation of PSMP to form aggregates. Sizes in the brackets are the area equivalent radius. The size distribution obtained from our measurements reflects the step in the red box.}
\label{mechanism}
\end{figure}

Conventional aggregation of particles is a multi-step process, including micro-floc growth stage and large floc growth stage \cite{BUBAKOVA2013540}. Primary particles (basic units) form micro-flocs (intermedium basic units) with a coagulant at the micro-floc growth stage. At the large floc growth stage, micro-flocs form large flocs by binding points of micro-flocs \cite{WANG20181183}. The same mechanism may describe asphaltene precipitation (Figure \ref{mechanism}). Nano-aggregates (basic units) form primary submicron particles (PSMP) (intermedium basic units) in the precipitation stage. The PSMPs form fractal flocs via further aggregation at the aggregation stage. Fractal aggregates are formed by aggregation of PSMP rather than directly incorporating asphaltene nano-aggregates. The size distribution discussed in this study is the particles larger than 0.2 $\mu m$ in radius that was detectable in our TIRF images.

Population balance model (PBM) will be used to describe the particle aggregation and to relate the size distribution of asphaltene aggregates with the properties of the precipitant. Smoluchowski equation for aggregation of primary units is \cite{elimelech2013particle}: 

\begin{equation}
\label{Sm}
\frac{dn_k}{dt}=\frac{1}{2} \sum_{i+j=k} K_{i,j} n_i n_j - n_k \sum_{i \ge 1} K_{i,k} n_i
\end{equation}
where $n_i$, $n_j$, and $n_k$ are the number concentration of particles with sizes i, j, and k, respectively. $t$ is time. $K_{i,j}$ is the collision kernel of aggregation, which can be estimated by Equation (\ref{K}): \cite{Maqbool2011}

\begin{equation}
\label{K}
K_{i,j} = \frac{2 R T}{3 \mu} \frac{(d_i + d_j)^2}{d_i d_j} \beta
\end{equation}
where R is the ideal gas constant, T is temperature, $\mu$ is viscosity, $d_i$ and $d_j$ are the diameters of particles with size i and j, respectively, and $\beta$ is collision efficiency, which can be estimated by Equation (\ref{beta}) \cite{Maqbool2011}:

\begin{equation}
\label{beta}
\beta \propto exp [- \frac{\psi_e}{k_B T (\delta_{asp} - \delta_{sol})^2}]
\end{equation}
where $\psi_e$ is the maximum energy barrier constant, $k_B$ is Boltzmann constant, $\delta_{asp}$ and $\delta_{sol}$ are Hildebrand solubility parameters of asphaltene and the paraffinic solvent, respectively. In the diffusive mixing process, $\delta_{sol}$ varies with time due to the change in the concentration of the paraffinic solvent, depending on the concentration of the paraffinic solvent in solution B and the diffusion coefficient \cite{ANGLE2006492,wang2003asphaltene}:     

\begin{equation}
\label{delta}
\delta_{sol} = \delta_{alk} \phi_{alk} + \delta_{tol} \phi_{tol} = \delta_{alk} \times \phi(t,l) + \delta_{tol} \times (1-\phi(t,l))
\end{equation}
where $\delta_{alk}$ and $\delta_{tol}$ represent Hildebrand solubility of the paraffinic solvent and toluene, respectively, $\phi_{alk}$ and $\phi_{tol}$ are volume fractions of the paraffinic solvent and toluene, respectively, and $\phi(t,l)$ is the volume fraction of the paraffinic solvent at a given time and position: \cite{meng2021microfluidic}

\begin{equation}
\label{phi}
\phi(t,l) = \phi_0 erfc (\frac{l}{2 \sqrt{D_{21}t}})
\end{equation}
where $l$ is the distance to the side channel, $\phi_0$ is the initial volume fraction of the paraffinic solvent, and $t$ is time.

In this study, the mean size ($r_{psmp}$) of the primary sub-micron particles (PSMP) (i.e., basic units) was 0.3 $\mu m$ in radius \cite{meng2021primary}. Asphaltene was destabilized and formed PSMP. It was assumed that all of the asphaltene PSMP were formed at the moment that the paraffinic solvent concentration was higher than the onset (i.e., the threshold of solvent concentration to promote the asphaltene precipitation as PSMP) upon destabilization kinetics was relatively quick compared with further aggregation of PSMP. The surface coverage of asphaltene particles ($SC$) observed in TIRF images was not affected by the aggregation of PSMP. Therefore, the initial quantity ($Q_p$) for the PBM was calculated based on $SC$ divided by the area of PSMP at the final state, as shown in Equation \ref{quantity}:

\begin{equation}
\label{quantity}
Q_p=SC/\pi r_{psmp}^2
\end{equation}
where $Q_p$ is the initial quantity of the PSMP.

Based on these assumptions and relations, the initial Smoluchowski equation can be rewritten as below: 

\begin{equation}
\label{Sm_sim}
\frac{dn_k}{dt}=\frac{8 R T}{3\mu} exp [- \frac{\psi_e}{k_B T (\delta_{asp} - (\delta_{alk} \times \phi(t,l) + \delta_{tol} \times (1-\phi(t,l)))^2}]\sum_{i+j=k} n_i n_j - n_k \sum_{i \ge 1} n_i
\end{equation}

The size distribution of the asphaltene particles at different times can be calculated by solving Equation (\ref{Sm_sim}). The asphaltene aggregates in the experimental results will be classified into three size bins. The number of equations to be solved can be narrowed down to three ordinary differential equations (ODEs), representing three bins for the particles sizes. The first bin represented PSMP ($\sim$ 0.3 $\mu m$), the second bin was for the particles with a size of 0.4 $\mu m$ to 0.6 $\mu m$, and the last one was for the particles with a size of 0.6 $\mu m$ to 0.8 $\mu m$. The particles larger than 0.8 $\mu m$ were not considered due to the quantity of them being less than 20\% of the total quantity in the experimental data. The parameters which depend on the paraffinic solvent were specified for each condition, and the equations were solved separately. The first part of Equation (\ref{Sm_sim}) on the right side of the equation is a function of time and did not depend on the number density of particles, and thus, an parameter B(t) was introduced as:

\begin{equation}
\label{B}
B(t)=\frac{8 R T}{3\mu} exp [- \frac{\psi_e}{k_B T (\delta_{asp} - (\delta_{alk} \times \phi(t,l) + \delta_{tol} \times (1-\phi(t,l)))^2}]
\end{equation}

Considering the birth and death of particles in each bin with aggregation of smaller particles and collisions with other particles to form larger particles, the three mentioned equations can be written as followed:

\begin{subequations}
\label{finaleqn}
\begin{align}
\frac{dn_1}{dt}=-B(t) n_1 (n_1 + n_2 + n_3 )\\
\frac{dn_2}{dt}=B(t) ( n_1 n_1 - n_2 (n_1 + n_2 + n_3 ))\\
\frac{dn_3}{dt}=B(t) ( n_1 n_2 - n_3 (n_1 + n_2 + n_3 ))
\end{align}
\end{subequations}

$n_1$, $n_2$, and $n_3$ refer to the concentration of the three bin sizes. The systems of ODEs were solved numerically using MATLAB (2021a) for each condition, and the results were compared with the experimental measurements. The solution strategy in MATLAB  was to use the standard solver ode45, which uses a six-stage, fifth order, Runge-Kutta method to solve the equations. The time step was 0.001 s, and for the initial condition, it was assumed that all of the particles were in the first bin (primary particles) at $t=0$. The equations were coded as a function in MATLAB, and ode45 was used to calculate the final size distribution. $\psi_e$ in PBM were estimated based on the group of $\phi_0^{pen}$ of 100 \% and applies to all of the conditions. The size distribution of asphaltene aggregates was modelled based on solving the PBM, collision kernels, and initial conditions.

\section{Results}

\subsection{Morphology of asphaltene particles precipitated induced by alkane and toluene mixture}

\begin{figure}[ht]
\centering
\includegraphics[width=1\columnwidth]{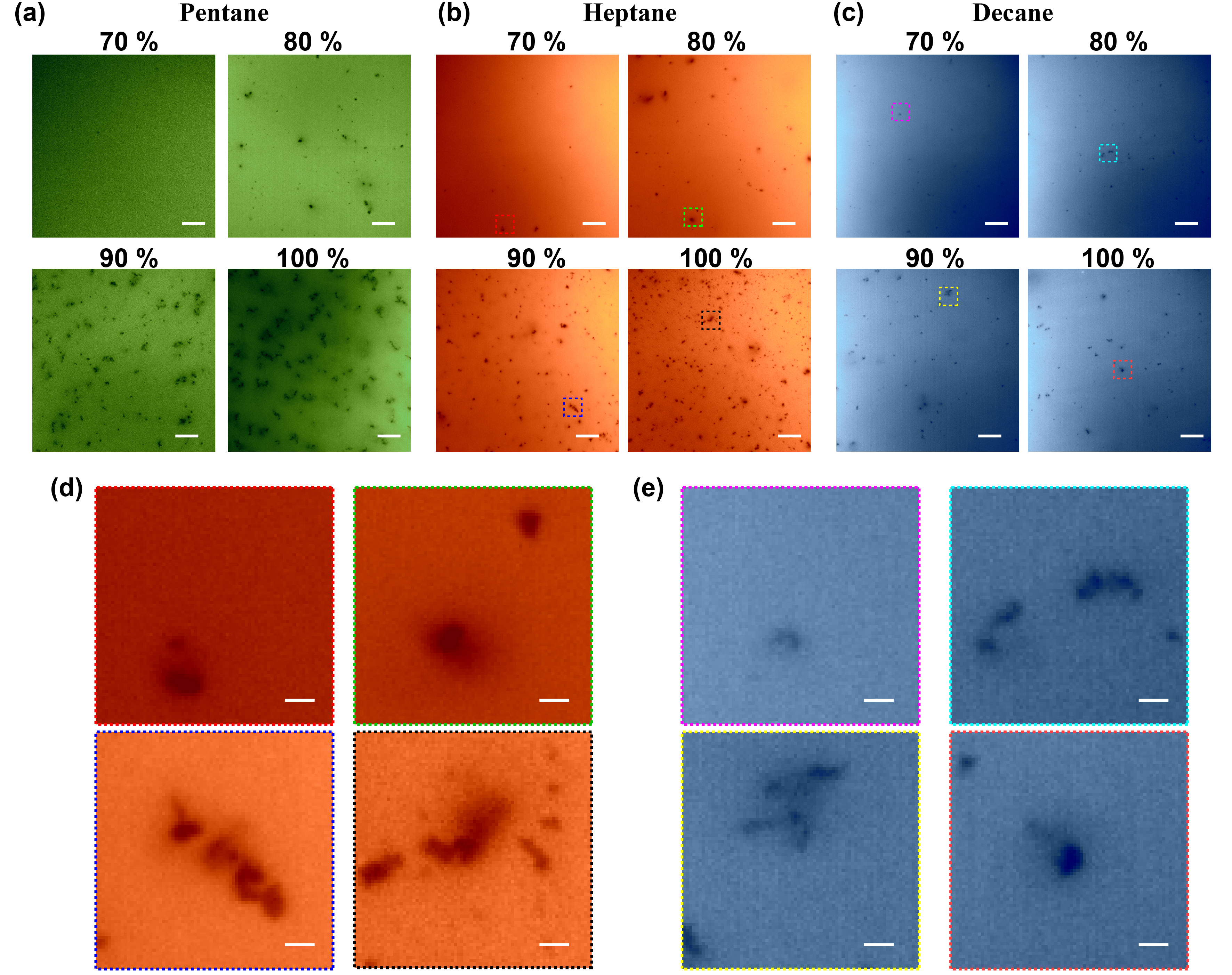}
\captionsetup{font={normal}}

\caption{TIRF images of asphaltene particles at the final state. Precipitation is induced by a) pentol, b) heptol, and c) dectol, and the corresponding zoomed-in images of d) heptol and e) dectol at locations with respective color boxes. The length of the scale bar is 10 $\mu m$ in (a-c), and is 1 $\mu m$ in (d-e). The images are false-colored.}
\label{alktol_morphology}
\end{figure}

Figure \ref{alktol_morphology}(a)\textendash (c) are the images of asphaltene particles precipitated in 12 types of precipitants, including pentol (pentane-toluene mixture), heptol (heptane-toluene mixture), and dectol (decane-toluene mixture). In the TIRF images, captured at the end of mixing \cite{meng2021primary}, the black dots represent the precipitated asphaltene particles as portrayed in Figure \ref{alktol_morphology}(a)\textendash (c).

The TIRF images of the precipitates show both individual particles (defined as PSMP) and fractal aggregates. The dark speckles in the fractal aggregates have a similar size to the individual particles. Individual particles and fractal aggregates are observed in pure heptane and decane, as shown in Figure \ref{alktol_morphology}(d)(e). The formation of PSMP is in good agreement with the previous study of pentane \cite{meng2021primary}. Furthermore, PSMP and fractal aggregates also form by mixing with a diluted alkane of $\phi_0$ with the concentration from 70\% to 90\%. The results show that the formation of the PSMP is ubiquitous in asphaltene precipitation induced by mixing with a paraffinic solvent, regardless of the type and concentration of the solvent. 

\begin{figure}[ht]
\centering
\includegraphics[width=\columnwidth]{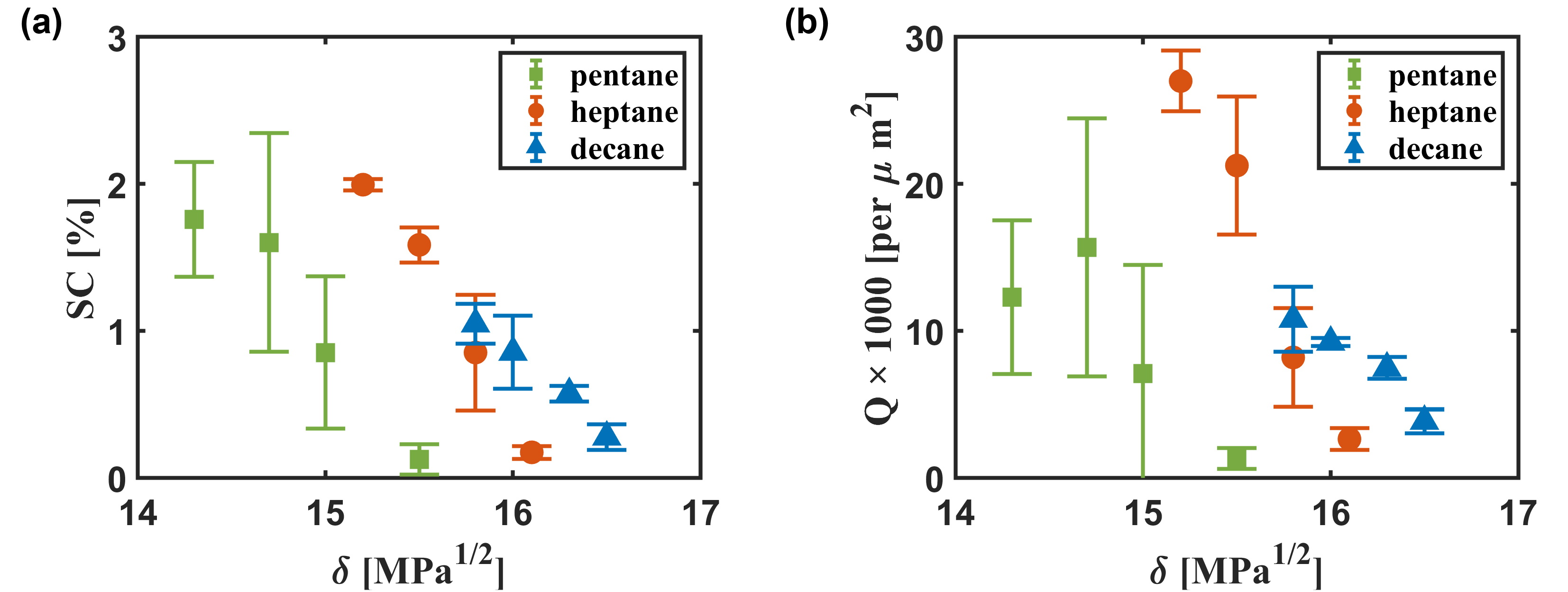}
\captionsetup{font={normal}}

\caption{a) Surface coverage ($SC$) and b) particle quantity ($Q$) in the unit area of the final state of asphaltene particles. Precipitation is induced by pentol, heptol, and dectol. For pentol, heptol, and dectol, Hildebrand solubility parameter ($\delta$) from high to low corresponds to $\phi_0$ ranging from 70\% to 100\%.}
\label{SCandQ_HSP}
\end{figure}

The surface coverage ($SC$) of asphaltene precipitates reflects the yield of the asphaltene during precipitation. The yield of asphaltene is the lowest in decane for the pure solvents, and the difference between pentane and heptane is not notable. For each type of solvent, the $SC$ of asphaltene particles increases with the initial alkane concentration ($\phi_0$). It means the yield of the asphaltene increases with $\phi_0$. For the same type of solvent, $\delta$ decreases with the increase of $\phi_0$, leading to increased $SC$ and the quantity ($Q$) of asphaltene particles in the final state. Figure \ref{SCandQ_HSP}(a)(b) show the quantitative analysis of $SC$ and $Q$ based on the TIRF optical images. Interestingly, to reach the same $SC$ or $Q$, the requirement of $\delta$ follows the trend of $\delta_{pen} < \delta_{hep} < \delta_{dec}$. In other words, $\delta$ of the precipitant is not the only dominated factor that determines $SC$ and $Q$ of asphaltene precipitates.

The same $SC$ of two groups does not necessarily correspond to similar $Q$. For example, $SC$ of 100\% of $\phi_0^{pen}$ is close to 100\% of $\phi_0^{hep}$. However, $Q$ of 100\% of $\phi_0^{pen}$ is about half of $Q$ from 100\% of $\phi_0^{hep}$. The reason is that aggregates are treated as one unit in $Q$ analysis. The $SC$ of one aggregate is higher than that of one PSMP. 

\begin{figure}
\centering
\includegraphics[width=1\columnwidth]{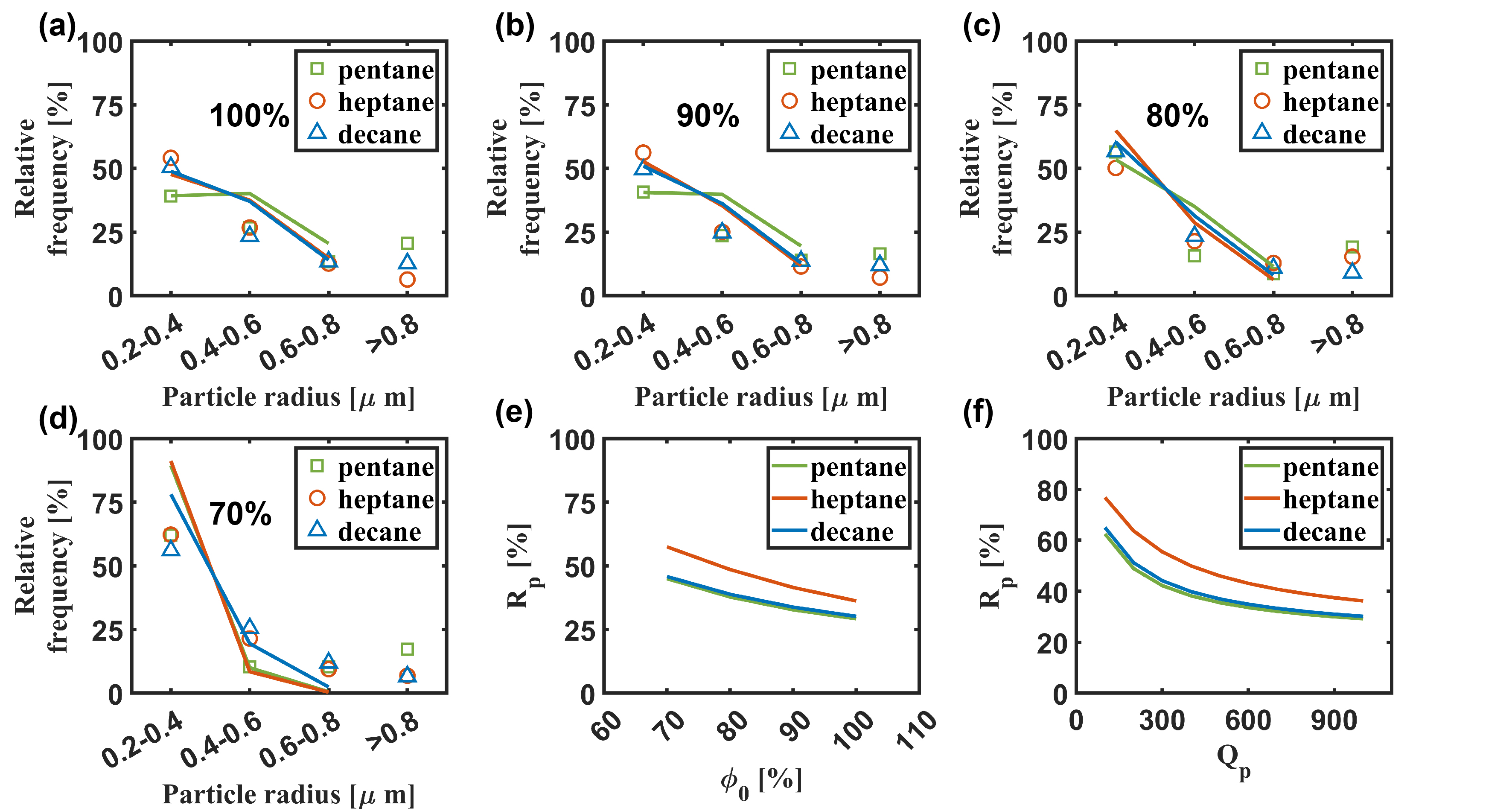}
\captionsetup{font={normal}}

\caption{Relative frequency of asphaltene particle size distribution with $\phi_0$ of a) 100\%, b) 90\%, c) 80\%, d) 70\%. The solid lines show the results predicted from PBM. e) Effect of n-alkane on $R_p$. The initial condition of the quantity of PSMP
is 1000. f) Effect of the initial quantity of PSMP on $R_p$. The initial condition of $\phi_0$ of the solvents is 100\%.
At a low quantity of PSMP region (i.e., $Q_p$ $<$ 300), the influence of initial condition on $R_p$
is very significant. But this dependence is not obvious at high quantity of PSMP region
(i.e., $Q_p$ $>$ 600).}
\label{rel_freq}
\end{figure}

Analysis of the size distribution of asphaltene particles shows that the number of PSMP is always the highest, followed by the aggregates with the size of 0.4 to 0.6 $\mu m$. The same trend was observed, regardless of the type of solvent and solvent concentration in the mixture. The $R_p$ is defined as the ratio between the number of the individually dispersed PSMP to $Q$. Notably, at $\phi_0$ of 100\% and 90\%, the $R_p$ of pentane is lower than heptane and decane, as shown in Figure \ref{rel_freq}(a)(b). But in pentane, more very large aggregates are formed in the range larger than 0.8 $\mu m$. At low $\phi_0$, $R_p$ of pentane, heptane, and decane are similar, as shown in Figure \ref{rel_freq}(c)(d). Nevertheless, large aggregates ($>$ 0.8 $\mu m$) form in pentane is more significant than in heptane and decane.

The size distribution of asphaltene particles estimated by PBM is shown in Figure \ref{rel_freq}(a)-(d). The simulated data agree well with the experimental data except for $\phi_0$ of 70\%. The reason for this error is that there are too few statistics (less than 20 particles) because the concentration of the paraffinic solvent is close to the onset, leading to the difference between the experimental and fitting results. Good agreement between the experimental data and prediction of PBM in particle size distribution suggests that the asphaltene growth kinetics from PSMPs to larger aggregates may be well explained by the aggregation of nano-colloids. The effects of different types of precipitants on the kinetics of asphaltene precipitation are described by varying the collision kernel and number density of asphaltene particles.

At a high solvent concentration of paraffinic solvent, PSMP yield is high, which gives rise to lower $R_p$ (Figure \ref{rel_freq}(e)\ref{rel_freq}(f)). The minimum of $R_p$ appears at pentane of $\phi_0^{pen}$ of 100 \%, which has the largest $SC$ (i.e., the largest quantity of PSMP). The influence of the initial condition of $Q_p$ on $R_p$ is significant at a small quantity of PSMP range (i.e., $Q_p$ $<$ 300), but it is not obvious at a large quantity of PSMP range (i.e., $Q_p$ $>$ 600). The collision frequency (i.e., the probability of particle collisions) dominates at a small particle quantity, while for a large particle quantity, the collision efficiency (i.e., the success rate of aggregation dominates $R_p$. The collision efficiency depends on the type of solvent, because of the difference of Hildebrand solubility parameter ($\delta$). Figure \ref{rel_freq}(e)(f) also indicates that $R_p$ of heptane is higher than pentane and decane, although our experimental data are not as distinct as the prediction from the model.

\subsection{Morphology of asphaltene particles precipitated induced by the mixture of heptane and decane }

\begin{figure}[ht]
\centering
\includegraphics[width=1\columnwidth]{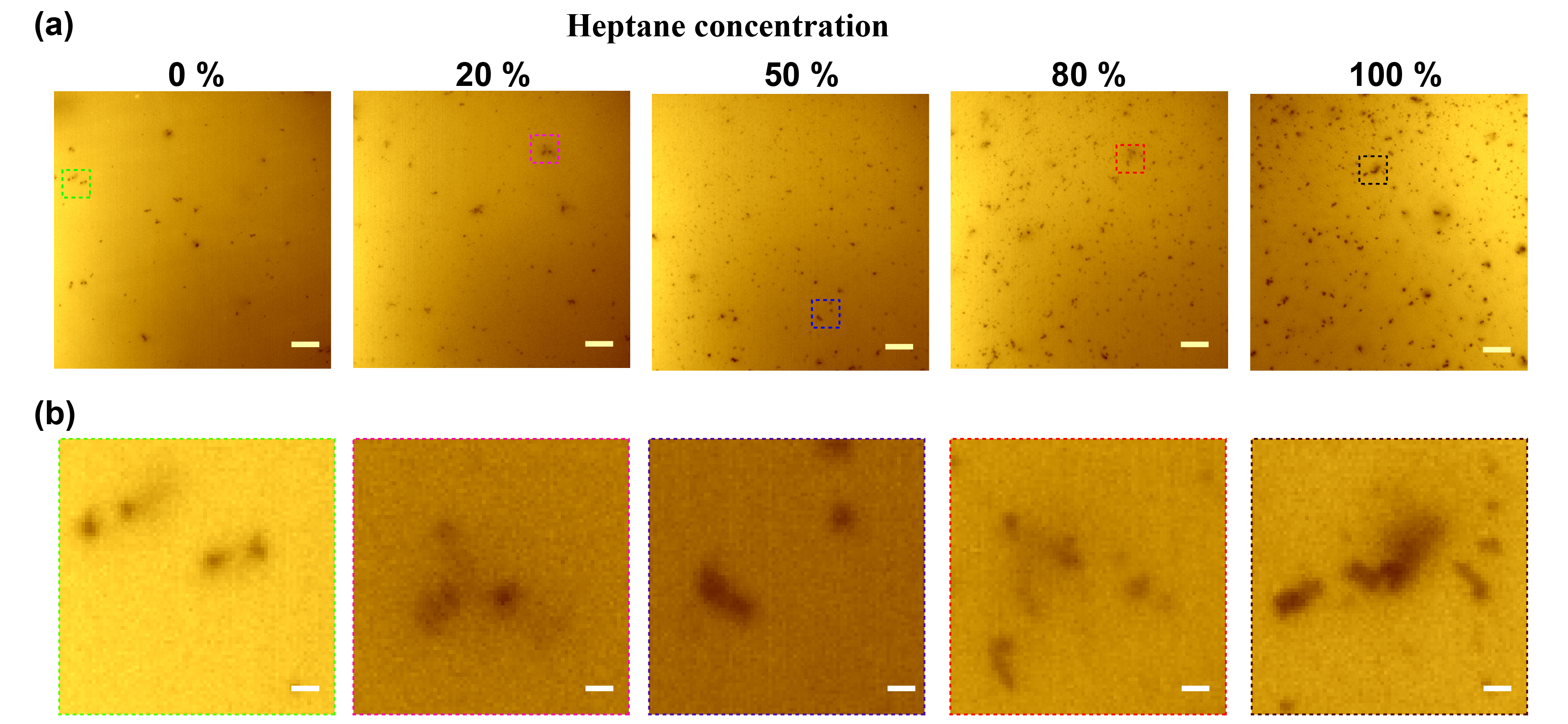}
\captionsetup{font={normal}}

\caption{a) TIRF images of asphaltene particles at the final state. The mixture of heptane and decane induces precipitation, and b) the corresponding zoomed-in TIRF images at locations with respective color boxes. The length of the scale bar is 10 $\mu m$ in (a), and is 1 $\mu m$ in (b). The images are false-colored. 
}
\label{hepdec_mix}
\end{figure}

Figure \ref{hepdec_mix}(a) shows the TIRF image of asphaltene particles precipitated in the mixture of heptane and decane with 0\% to 100\% heptane. It is found that, with the increase of heptane concentration, more asphaltene particles are formed. Figure \ref{hepdec_mix}(b) shows that the formation of both PSMP and aggregates, further illustrating the ubiquitous presence of the PSMP in the mixture of paraffinic solvents.

\begin{figure}
\centering
\includegraphics[width=1\columnwidth]{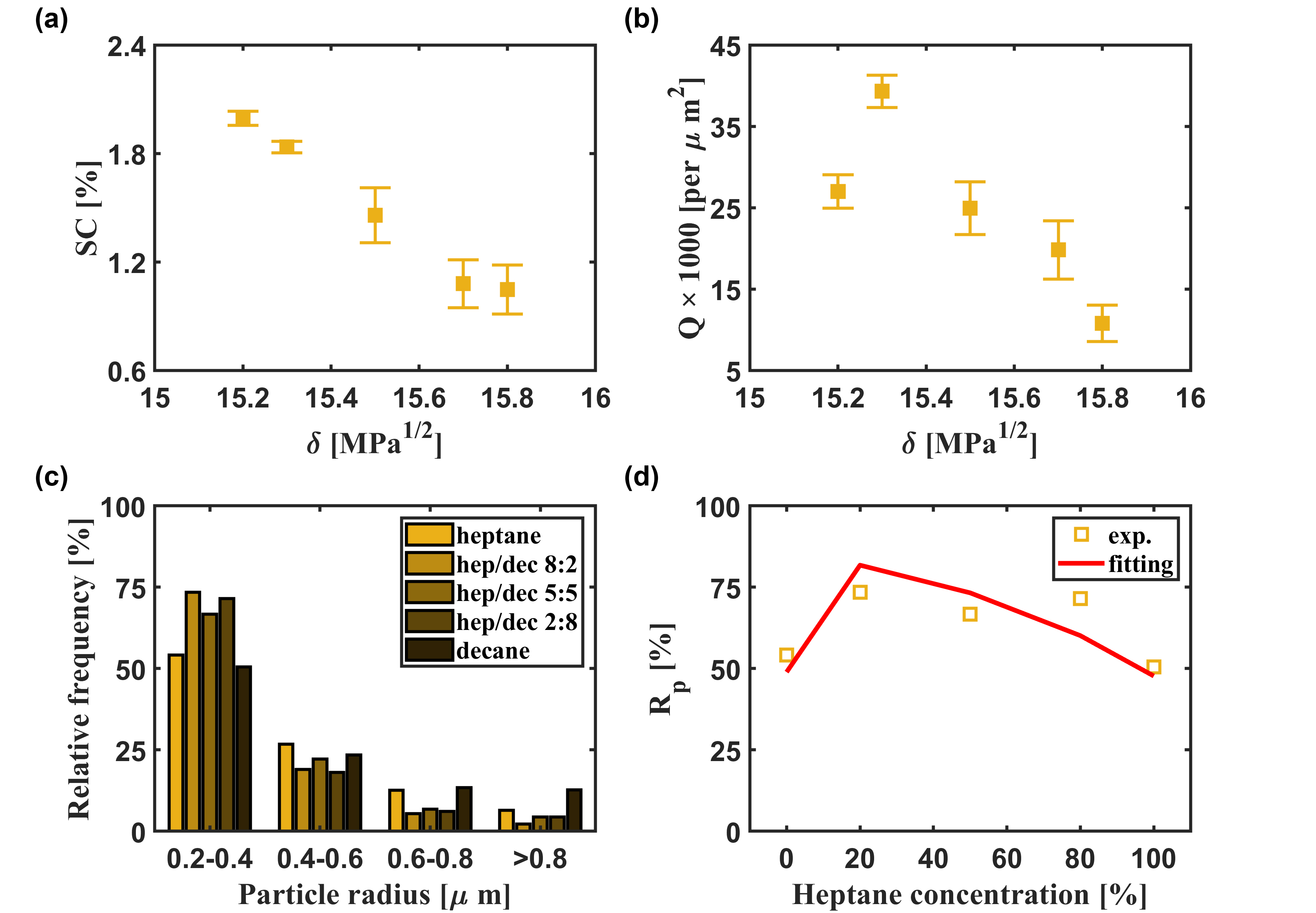}
\captionsetup{font={normal}}

\caption{a) Surface coverage ($SC$) and b) particle quantity ($Q$) in the unit area of asphaltene particles at the final state. Precipitation is induced by the mixture of heptane and decane. Hildebrand solubility parameter ($\delta$) from high to low corresponds to heptane concentration from 0\% to 100\%. c) Relative frequency of asphaltene particle size distribution for different heptane and decane mixtures. d) Comparison of $R_p$ of the experimental and prediction results from PBM.
}
\label{hepdec_fre}
\end{figure}

Consistent with the results of asphaltene precipitation induced by pentol, heptol, and dectol, both $SC$ and $Q$ decrease with the increase in $\delta$ of the mixture of heptane and decane as shown in Figure \ref{hepdec_fre}(a)(b). Also, PSMP is the most, followed by aggregates in the next size bin from 0.4 to 0.6 $\mu m$, which is the same as pentol, heptol, and dectol. It is important to note that the $R_p$ of the three mixture groups is higher than the pure heptane and decane. However, Figure \ref{hepdec_fre}(c) revealed that the ratio of aggregates ranging from 0.4 to 0.6 $\mu m$ of the mixtures has the consistently opposite relationship with the PSMP.


Interestingly, Figure \ref{hepdec_fre}(c) shows that asphaltene particles precipitated in the mixtures of heptane and decane have higher $R_p$ compared with pure heptane or decane. This attributes to the different diffusion coefficients of heptane and decane to toluene. Heptane diffuses and induces the initial formation of PSMP faster. Decane follows up and dominates the subsequent aggregation of the PSMP. As shown in Equation (\ref{beta}), collision efficiency in decane is lower than heptane, so that the probability of the aggregates is lower than pure heptane. The PBM results are consistent, and show induces larger $R_p$ (Figure \ref{hepdec_fre}d).

\subsection{Morphology of asphaltene particles precipitated in bitumen}

Compared with our model oil of 17 g/L asphaltene in toluene in the present work, the chemical composition of the natural bitumen is more complicated. More than asphaltene, bitumen contains saturates, aromatic, and resins (SAR) \cite{gray2015upgrading}. In this investigation, bitumen was also studied to examine the influence of SAR.

\begin{figure}[ht]
\centering
\includegraphics[width=1\columnwidth]{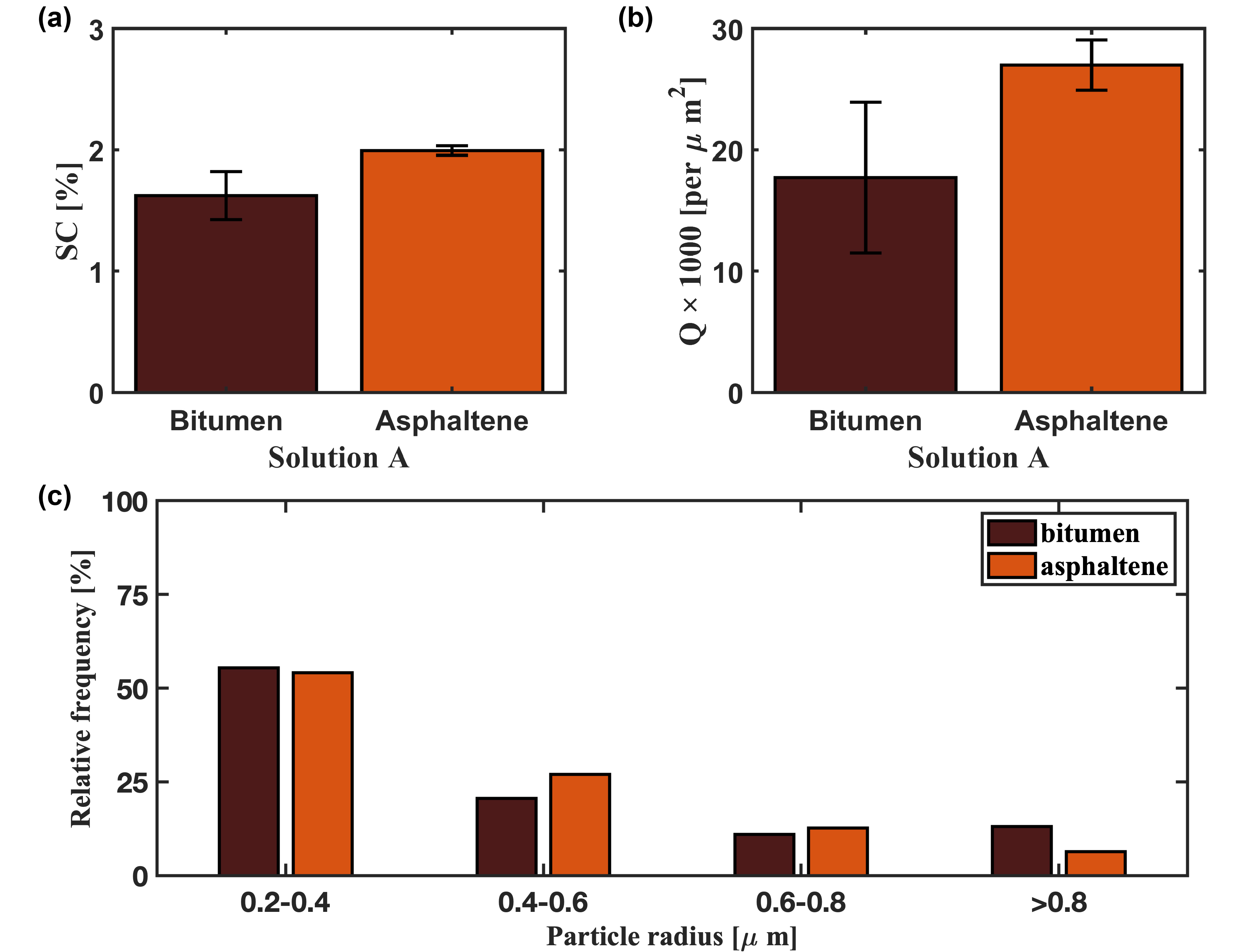}
\captionsetup{font={normal}}

\caption{a) Surface coverage ($SC$) and b) particle quantity ($Q$) in the unit area of the final state of asphaltene particles precipitate from bitumen and asphaltene solution. Precipitation is induced by heptane. c) Comparison of the relative frequency of asphaltene particle size distribution for asphaltene and bitumen solution.
}
\label{bitumen}
\end{figure}

The TIRF optical images are shown in Figure S1. PSMP also exists in the precipitate form of asphaltene particles from bitumen. As shown in Figure \ref{bitumen}(a) and \ref{bitumen}(b), $SC$ and $Q$ of asphaltene particles from bitumen solution and model oil system are similar. The minimal difference may be caused by the uncertainty of asphaltene concentration in bitumen. The $R_p$ of bitumen and asphaltene solution are also close. SAR does have significant impact on the asphaltene precipitation, but it just did not observed by our technique. One contribution for SAR is the solvency effect that they have a similar effect as good solvents (e.g. AR) and poor solvent (e.g. S). Also the toluene in solution A is significant which may have an even better solvency effect than AR fractions. Our technique does not have the sensitivity to distinguish the difference when the particles have grown to sub-microns. 


\subsection{Morphology of asphaltene particles precipitated in the solution with inhibitor}

\begin{figure}
\centering
\includegraphics[width=1\columnwidth]{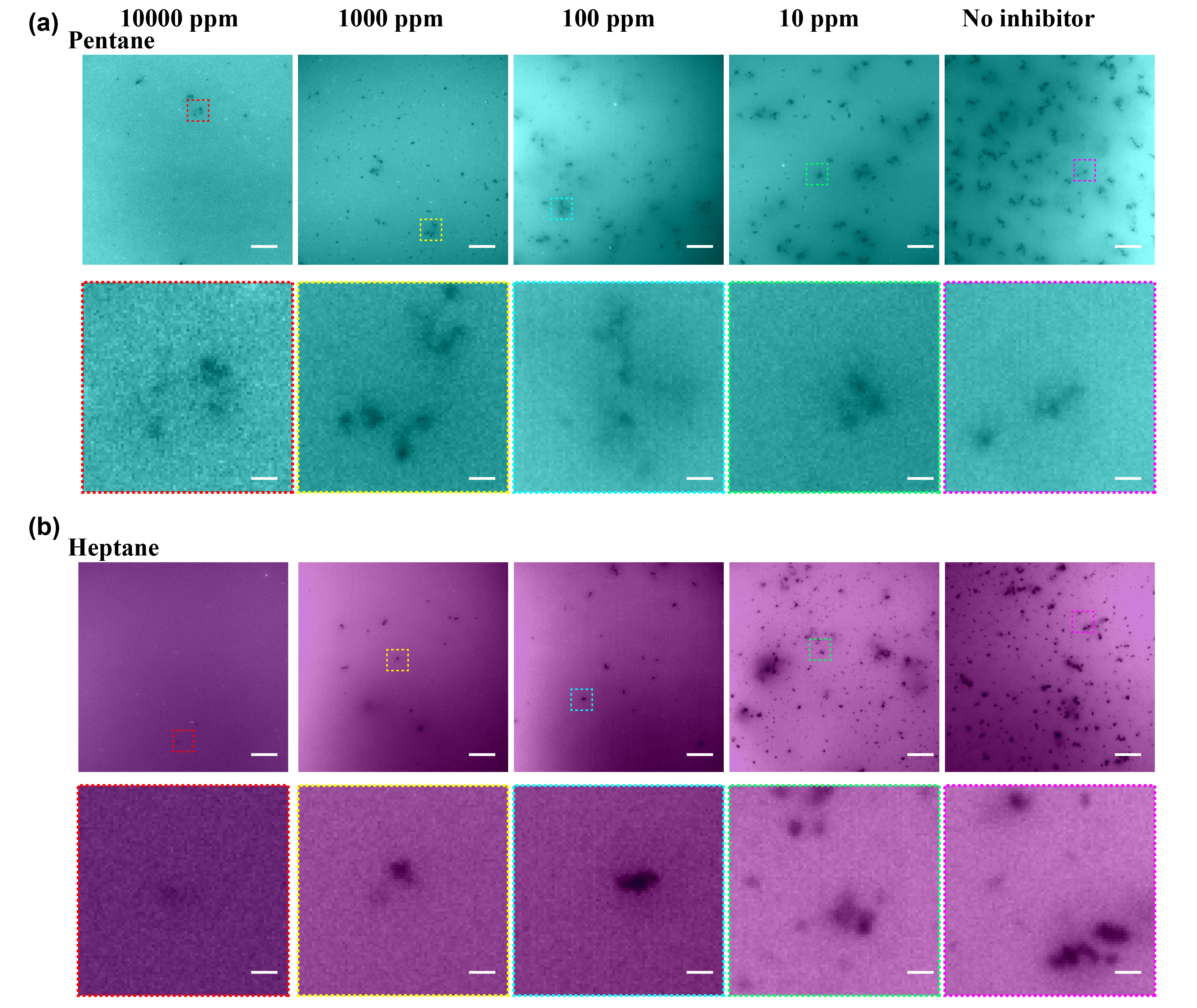}
\captionsetup{font={normal}}

\caption{TIRF images of asphaltene particles at the final state with the addition of inhibitor from 0 to 10000 ppm. Precipitation is induced by a) pentane and b) heptane. The bottom rows are zoomed-in images of PSMP at locations with respective color boxes. Note: The images have been false-colored. The length of the scale bar is 10 $\mu m$, and is 1 $\mu m$ in the zoomed-in images. The images are false-colored.
}
\label{inhibitor_opt}
\end{figure}

Chemicals that can prohibit the precipitation of asphaltene are called inhibitors \cite{subodhsen1999,balestrin2019}. In this study, we keep the concentration of the inhibitor of nonylphenol in the asphaltene solution and precipitant the same. Figure \ref{inhibitor_opt}(a)(b) are the TIRF optical images of asphaltene particles precipitate with the addition of the inhibitor in pentane and heptane at the concentration of the inhibitor from 0, 10, 100, 1000, to 10000 $ppm$. It is worth noting that some of the asphaltene particles appear as white particles in the TIRF optical images, indicating higher fluorescence intensity from the particles than from the surrounding medium. This may be due to the quenching effect of asphaltene particles being reduced by the inhibitor. The zoomed-in images show the formation of PSMP. As inhibitor concentration increases, fewer asphaltene particles are observed from TIRF images (Figure \ref{inhibitor_opt}).

\begin{figure}
\centering
\includegraphics[width=1\columnwidth]{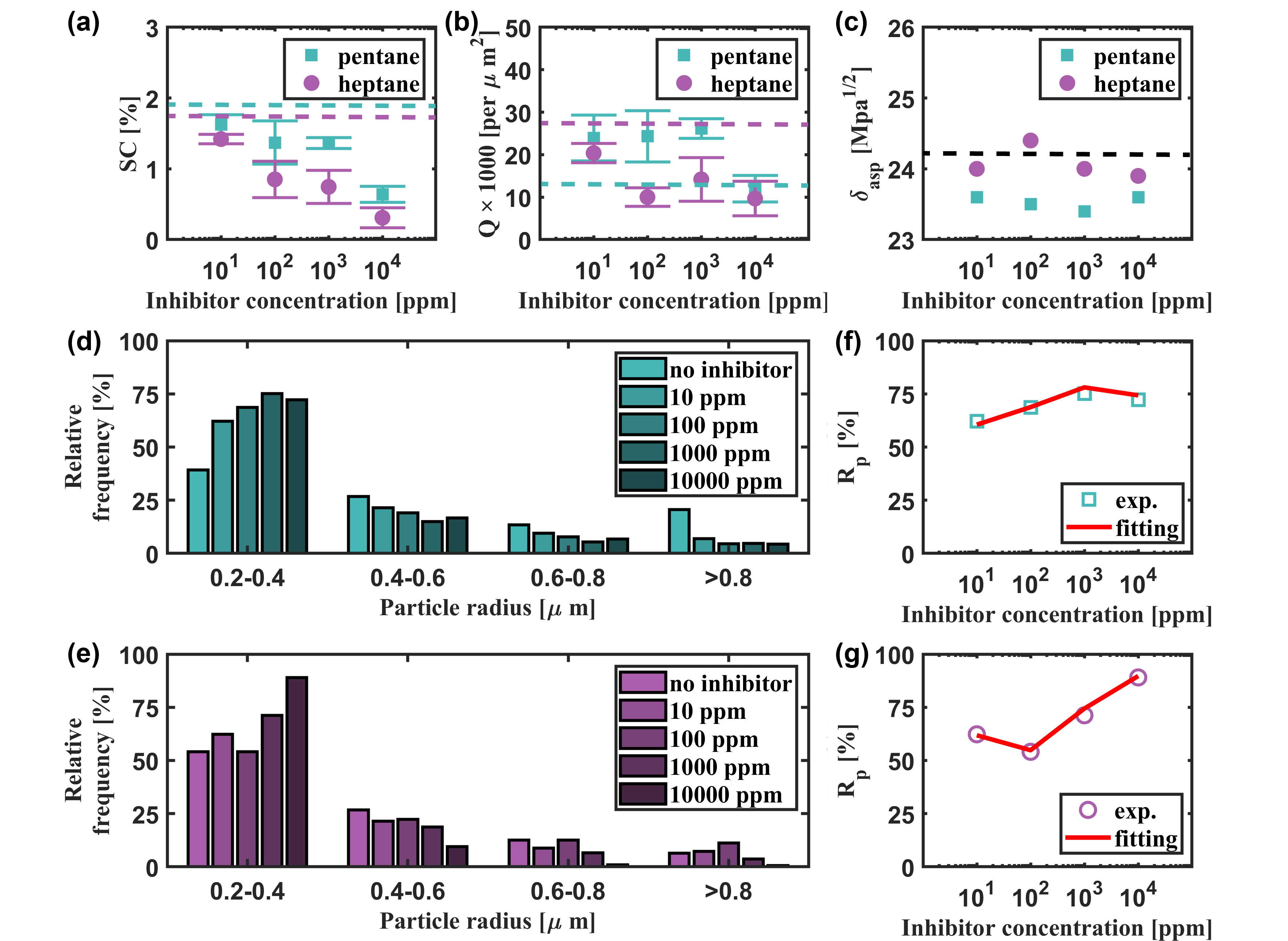}
\captionsetup{font={normal}}

\caption{a) Surface coverage ($SC$) and b) particle quantity ($Q$) as a function of the concentration of inhibitor. Cyan and purple dotted lines $SC$ and $Q$ of asphaltene particles precipitated in pentane and heptane without inhibitor, respectively. c) $\delta_{asp}$ is varied in the PBM model to fit the experimental data. The black dotted line represents $\delta_{asp}$ of inhibitor-free asphaltene. Influence of inhibitor on the particle size distribution of asphaltene particles precipitated in d) pentane and e) heptane. Comparison of $R_p$ of the experimental and prediction results from PBM in f) pentane and g) heptane.}
\label{inhibitor}
\end{figure}

In literature, 10000 $ppm$ nonylphenol can reduce asphaltene precipitation \cite{balestrin2019}. We found that the inhibiting effect depends on the types of paraffinic solvent. As shown in Figure \ref{inhibitor}(a), the $SC$ of asphaltene particles decreases with the addition of nonylphenol. As the concentration of nonylphenol increases, the lower surface is covered by asphaltene particles for both pentane and heptane. However, for heptane, both formations of PSMP and fractal aggregates are inhibited, which is reflected in the corresponding reduction of $SC$ and $Q$, as shown in Figure \ref{inhibitor}(a)(b). In the case of pentane, the effect of nonylphenol on inhibition is mainly to the aggregation of PSMP. Therefore, while decreasing $SC$, $Q$ is increasing.

Figure \ref{inhibitor}(d)(e) show the two highest peaks are of PSMP and aggregates from 0.4 to 0.6 $\mu m$ with the addition of the inhibitor in pentane and heptane. In case of pentane, as nonylphenol concentration increases, the inhibition effect of aggregation of PSMP becomes stronger, resulting in $R_p$ increases and less fractal aggregates forming. For heptane, adding nonylphenol from 0 to 100 $ppm$ leads to a similar decrease in the formation of PSMP and aggregation. The size distribution does not change significantly. The addition of nonylphenol from 100 to 10000 $ppm$ inhibits relatively more aggregation of PSMP than the formation of PSMP. For the change from 100 to 10000 $ppm$, heptane shows similar results to pentane.

Asphaltene molecules have been shown to form nano-aggregates in toluene at low concentrations ($\sim$ 0.1 $g/L$) \cite{Mostowfi2009,Betancourt2009}. The concentration of asphaltene 17 $g/L$ is used in this study where the majority motif of asphaltene is nano-aggregates. In colloidal theory, the extended structure of the asphaltene colloids collapses in precipitant, so van der Waals attraction leads to the aggregation of asphaltene nano-aggregates and appears as a new phase formation when the asphaltene aggregates grow large enough \cite{gray2011,wang2010}. 

The inhibitor molecules may adsorb on the surface of asphaltene nano-aggregates and becomes part of the extended structure around polyaromatic cores \cite{Rogel2011}. The adsorbed inhibitor molecules enhance the steric repulsion between asphaltene nano-aggregates \cite{Rogel2011}, leading to lower attraction among nano-aggregates in precipitants. Thus, adding the inhibitor, whose $\delta$ is 19.3 $MPa^{1/2}$ \cite{barton2017crc}, decreases the solubility parameter of asphaltene ($\delta_{asp}$). With decreasing $\delta_{asp}$, PBM model can fit the $R_p$ at different inhibitor concentrations, as shown in Figure \ref{inhibitor}(f)(g). The exact values of $\delta_{asp}$ for the PBM fitting are shown in Figure \ref{inhibitor}(c). 

The molar mass of asphaltene nano-aggregates is approximately 1,000 to 30,000 $g/mol$ \cite{Acevedo2005,Yarranton2000,Barrera2013}. Therefore, nano-aggregate concentration in our system is 5.7 $\times$ $10^{-4}$ $mol/L$ to 1.7 $\times$ $10^{-2}$ $mol/L$ based on 17 $g/L$ asphaltene in toluene. For 10,000 $ppm$ inhibitor, the concentration is 4.3 $\times$ $10^{-2}$ $mol/L$, which is higher than the concentration of asphaltene nano-aggregates. On average, each nano-aggregate at least adsorbs one nonylphenol molecule, resulting in the decrease of $\delta_{asp}$ for all asphaltene. The estimated value of $\delta_{asp}$ is lower than 24 $MPa^{1/2}$, which is smaller than the pure asphaltene (i.e., 24.2 $MPa^{1/2}$.) Thus, nano-aggregates are less likely to grow large enough to PSMP as a new phase and PSMP is less likely to grow larger via further aggregation, resulting in $SC$ becomes to a third and $R_p$ doubles in both pentane and heptane cases.

In the case of  10 to 1,000 $ppm$ of inhibitor on average, some PSMP may have none or a minimal number of inhibitor molecules. $\delta_{asp}$ of the inhibitor-free asphaltene is not changed, and they can aggregate to form fractal aggregates as normal. The quantity of the inhibitor-free asphaltene particles increases with the decrease of the inhibitor concentration. Therefore, the $SC$ decreases, and $R_p$ increases accordingly with the increase of inhibitor concentration. 

However, for inhibitor concentration from 0 to 1000 $ppm$, $SC$ decreases slightly from 1.7\% to 1.4\%, but $R_p$ increases from 40\% to 80\% in pentane. This indicates that the inhibition in pentane is mainly on the aggregation of PSMP because the $\delta_{pen}$ (i.e., 14.3 $MPa^{1/2}$) is lower than $\delta_{hep}$ (i.e., 15.2 $MPa^{1/2}$). Although $\delta_{asp}$ decreases from 24.2 $MPa^{1/2}$ to $\sim$ 23.5 $MPa^{1/2}$ (Figure \ref{inhibitor}(c)), pentane is still strong enough to provide sufficient collision frequency for nano-aggregate to grow to PSMP after adding inhibitor. However, as the particle grows, the number of inhibitor molecules in one asphaltene particles increases, resulting in the further decrease of $\delta_{asp}$. The corresponding increase of steric repulsion results in PSMP not being aggregated. Then, pentane is not adequately strong for further aggregation. Therefore, most of the particles exist as PSMP, displayed as $R_p$ is approximately doubled with the addition of inhibitors. 

\section{Further discussion: Solvency influence on the yield of asphaltene particles}

Diffusion coefficient ($D_{21}$), collision frequency ($\alpha_{i,j}$), and efficiency ($\beta_{i,j}$) determine both yield and the particle size distribution (in PBM model). In addition, diffusion coefficients influence the yield via the mixing time.

Even though it is still controversial whether the asphaltene precipitation is a phase change process or a colloidal growth process, most works \cite{gray2011,wang2010,YUDIN1998297,ENAYAT2020116250} agree that larger absolute value of the difference of $\delta$ between asphaltene and precipitant enhances asphaltene precipitation. Hildebrand solubility parameter difference between asphaltene and pentane is larger than heptane and decane ($\delta_{asp} - \delta_{pen}$ > $\delta_{asp} - \delta_{hep}$ > $\delta_{asp} - \delta_{dec}$) \cite{Haji2013}. Larger difference of $\delta$ increases the collision efficiency of asphaltene nano-aggregates due to the decreasing of steric repulsion. Therefore, the rate of asphaltene precipitation in terms of $SC$ in pentane is higher than heptane and decane. 

However, we found that precipitants with the same $\delta$ do not necessarily lead to the same levels of $SC$ and $Q$. Specifically, $\delta$ for precipitants of $\phi_0^{pen}$ of 70 \% is the same as that of $\phi_0^{hep}$ of 90 \%. But both $SC$ and $Q$ of $\phi_0^{hep}$ of the latter are 15 times higher than those of the former. Our results suggest that apart from $\delta$ (thermodynamic aspect), the mixing process (hydrodynamic aspect) of asphaltene solution and the precipitant may play a role in asphaltene precipitation.

The diffusive mixing process in our quasi-2D micro channel allows for quantitative analysis of the effects from diffusion coefficients. Our previous work \cite{meng2021primary, meng2021microfluidic} showed that in the chamber with the same dimensions it took about tens seconds for the chemical composition to transit from the asphaltene solution to pure the paraffinic solvent supplied from the side chamber. At a given time in the mixing process, the ratio between precipitant to asphaltene ($R$) increases (Figure \ref{chamber}c) along the concentration gradient from the asphaltene solution towards the paraffinic solvent. Asphaltene precipitation takes place at the location where the ratio $R$ is above the onset. For a given location, the ratio $R$ gradually increases with time and reaches the onset, from which moment the asphaltene precipitation begins. Eventually, the mixture containing asphaltene in the location is nearly replaced by the precipitant, and the precipitation finishes due to the lack of the source of asphaltene. Mixing retention time ($t_R$) is defined as the time duration between the start of asphaltene precipitation and the end. For a higher diffusion coefficient(i.e., pentane), the retention time ($t_{R}$) of mixing is shorter. Therefore, mixing retention time of pentane ($t_{Rpen}$) is shorter than heptane ($t_{Rhep}$), and $t_{Rhep}$ is shorter than decane ($t_{Rdec}$).

The final yield of asphaltene is the result of the coupled effects from the precipitation rate of asphaltene and the mixing retention time. Although $\delta$ follows $\delta_{asp} - \delta_{pen}$ > $\delta_{asp} - \delta_{hep}$, the mixing retention time follows $t_{Rpen}$ < $t_{Rhep}$ due to $D_{pt}$ > $D_{ht}$. Consequently, the $SC$ of pentane and heptane are similar. On the other hand, the precipitation rate of asphaltene is low in decane, although the mixing retention time is longer than $t_{Rpen}$ and $t_{Rhep}$, $SC$ from decane is still lower than from pentane and heptane. 

For pentol, heptol, and dectol, the diffusion coefficients of alkanes with respect to initial $\phi_0$ vary in a small range (Figure \ref{HSP}c). Therefore, the effect from the difference of diffusion coefficient is negligible, compared to that from the variance of $\delta$ at different $\phi_0$. Both $SC$ and $Q$ increase with the decrease of $\delta$, as observed in our results.

\section{Conclusions}
23 types of precipitants are compared to induce asphaltene precipitation under diffusive mixing in a quasi-2D microfluidic chamber in this work. The precipitated asphaltene particles are captured by TIRF to exhibit a high spatial resolution. The formation of PSMP (i.e., size range from 0.2 to 0.4 $\mu m$ in radius) is ubiquitous in the diffusive mixing quasi-2D microfluidic chamber. The formation of PSMP is independent of the type and concentration of solvents, the presence of inhibitors, and the asphaltene solution is model oil or bitumen. However, for the size distribution, the mixture of heptane and decane produces the highest $R_p$ in the diffusive mixing process. The yield of asphaltene increases with the decrease of the Hildebrand solubility parameter, also influenced by the diffusion coefficient. Both the yield of the precipitated asphaltene and the PSMP ratio can be tuned by changing the type and composition of the precipitant. Population balance model (PBM) is developed and validated by the in-house experimental data in terms of the size distribution of the asphaltene particles. The results of $R_p$ from the PBM are in good agreement with the experimental results at different solvent conditions. In addition to the solvents, the inhibitor can prohibit both the formation and aggregation of PSMP in the diffusive mixing system. The experimental data of adding inhibitors can be fitted by adjusting $\delta_{asp}$ in PBM, which suggests that the prohibition effect of inhibitors may be caused by changing the structure of asphaltene aggregates. Excellent agreement of fitting shows that the parameters in PBM are likely to be reasonable. PBM provides the foundation for modelling real processes.

These findings may help to control the asphaltene amount and morphology to enhance the separation of bitumen from oilsands or the stability of heavy oil in transport.

\section*{Declaration of Competing Interests}
The authors declare that they have no known competing financial interests or personal relationships that could have appeared to influence the work reported in this paper.

\section*{Acknowledgement}
The authors acknowledge the funding support from the Institute for Oil Sands Innovation (IOSI) (project number IOSI 2018–03), from the Natural Science and Engineering Research Council of Canada (NSERC)-Collaborative Research and Development Grants, and from the Canada Research Chair Program and from Canada Foundation for Innovation, John R. Evans Leaders Fund. The authors are grateful for technical support from IOSI lab, particularly from Lisa Brandt and Brittany MacKinnon. We are also grateful for the technical support of Dr. Stephen Ogg and Greg Plummer at the Cell Imaging Center of Faculty of Medicine \& Dentistry's core microscopy facility and our industry steward Dr. Sepideh Mortazavi Manesh for fruitful discussion.

\section*{Nomenclature}
$\alpha_{i,j}$ = Collision frequency ($m^3/mol \cdot s$)

$\beta_{i,j}$ = Collision efficiency

$c_{asp}$ = concentration of asphaltene

$c_{pre}$ = concentration of precipitant

$c_{tol}$ = concentration of toluene

$\delta$ = Hildebrand solubility parameter ($MPa^{1/2}$)

$\delta_{asp}$ = Hildebrand solubility parameter of asphaltene ($MPa^{1/2}$)

$\delta_{dec}$ = Hildebrand solubility parameter of decane ($MPa^{1/2}$)

$\delta_{hep}$ = Hildebrand solubility parameter of heptane ($MPa^{1/2}$)

$\delta_i$ = Hildebrand solubility parameter of component i ($MPa^{1/2}$)

$\delta_{mixture}$ = Hildebrand solubility parameter of mixture ($MPa^{1/2}$)

$\delta_{pen}$ = Hildebrand solubility parameter of pentane ($MPa^{1/2}$)

$\delta_{tol}$ = Hildebrand solubility parameter of toluene ($MPa^{1/2}$)

$\Delta \rho$ = Density difference between alkanes and toluene ($kg/m^3$)

$d_i$ = Diameter of particle with size i

$d_j$ = Diameter of particle with size j

$D_{21}$ = Diffusion coefficient of alkane to toluene ($m^2/s$)

$D_{dt}$ = Diffusion coefficient of decane to toluene ($m^2/s$)

$D_{ht}$ = Diffusion coefficient of heptane to toluene ($m^2/s$)

$D_{pt}$ = Diffusion coefficient of pentane to toluene ($m^2/s$)

$g$ = Gravity acceleration constant ($m/s^2$)

$h$ = Height of the quasi-2D channel ($m$)

$k_B$ = Boltzmann constant ($J/K$)

$K_{i,j}$ = Collision kernels of particles with size $i$, $j$ ($m^3/mol \cdot s$)

$K_{i,k}$ = Collision kernels of particles with size $i$, $k$ ($m^3/mol \cdot s$)

$\mu$ = Viscosity ($m \cdot Pa$)

$m_p$ = Quantity of primary sub-micron particle in one aggregate

$n_i$ =  Units of aggregates with size $i$

$n_j$ =  Units of aggregates with size $j$

$n_q$ = Units of aggregates containing $q$ primary sub-micron particles. 

$n_o$ = Total units of primary sub-micron particle

$N$ = Maximum primary sub-micron particles that one aggregate contains 

$\phi_i$ = Volume fraction of component i (\%)

$\psi_e$ = Maximum energy barrier constant ($J \cdot MPa^{1/2}$)

$\phi(t,l)$ = Concentration of n-alkane at the given time and position 

$\phi^{alk}$ = Concentration of n-alkane (\%)

$\phi^{tol}$ = Concentration of toluene (\%)

$\phi_0$ = Initial concentration of alkane in solution B (\%)

$\phi_0^{dec}$ = Initial concentration of decane in solution B (\%)

$\phi_0^{hep}$ = Initial concentration of heptane in solution B (\%)

$\phi_0^{pen}$ = Initial concentration of pentane in solution B (\%)

PSMP = Primary sub-micron particle

PFT = Paraffinic froth treatment

$Q$ = Particle quantity in the unit surface area

$Q_p$ = Quantity of primary sub-micron particles at initial condition

$r_psmp$ = Radius of the primary sub-micron particle ($\mu m$)

$R$ = Ideal gas constant ($J$ $\cdot$ $K^{-1}$ $mol^{-1}$)

$Ra$ = Rayleigh number

$R_p$ = Ratio between primary sub-micron particle to the total quantity of asphaltene particles

S/B ratio = Solvent to bitumen ratio

$SC$ = Surface coverage (\%)

$t$ = Time ($s$)

$t_{onset}$ = Moment that precipitant concentration reaches onset

$t_{final}$ = Moment that asphaltene precipitation stops

$t_{Rdec}$ = Mixing retention time of decane and asphaltene solution ($s$)

$t_{Rhep}$ = Mixing retention time of heptane and asphaltene solution ($s$)

$t_{Rpen}$ = Mixing retention time of pentane and asphaltene solution ($s$)

$T$ = Temperature ($K$)

TIRF = Total internal reflection fluorescence microscope

\bibliography{main}

\begin{thebibliography}{10}
\expandafter\ifx\csname url\endcsname\relax
  \def\url#1{\texttt{#1}}\fi
\expandafter\ifx\csname urlprefix\endcsname\relax\def\urlprefix{URL }\fi
\expandafter\ifx\csname href\endcsname\relax
  \def\href#1#2{#2} \def\path#1{#1}\fi

\bibitem{gray2015upgrading}
M.~R. Gray, Upgrading oilsands bitumen and heavy oil, University of Alberta,
  2015.

\bibitem{chen2021ex}
P.~Chen, J.~N. Metz, A.~S. Gross, S.~E. Smith, S.~P. Rucker, N.~Yao, Y.~Zhang,
  \href{https://doi.org/10.1021/acs.energyfuels.1c02487}{Ex situ and in situ
  thermal transformations of m-50 pitch revealed by non-contact atomic force
  microscopy}, Energy \& Fuels (2021).
\newblock \href
  {http://arxiv.org/abs/https://doi.org/10.1021/acs.energyfuels.1c02487}
  {\path{arXiv:https://doi.org/10.1021/acs.energyfuels.1c02487}}.
\newline\urlprefix\url{https://doi.org/10.1021/acs.energyfuels.1c02487}

\bibitem{xu2018}
Y.~Xu, \href{https://doi.org/10.1021/acs.energyfuels.7b03013}{Asphaltene
  precipitation in paraffinic froth treatment: Effects of solvent and
  temperature}, Energy \& Fuels 32~(3) (2018) 2801--2810.
\newblock \href
  {http://arxiv.org/abs/https://doi.org/10.1021/acs.energyfuels.7b03013}
  {\path{arXiv:https://doi.org/10.1021/acs.energyfuels.7b03013}}, \href
  {https://doi.org/10.1021/acs.energyfuels.7b03013}
  {\path{doi:10.1021/acs.energyfuels.7b03013}}.
\newline\urlprefix\url{https://doi.org/10.1021/acs.energyfuels.7b03013}

\bibitem{Joshi2001}
N.~B. Joshi, O.~C. Mullins, A.~Jamaluddin, J.~Creek, J.~McFadden,
  \href{https://doi.org/10.1021/ef010047l}{Asphaltene precipitation from live
  crude oil}, Energy \& Fuels 15~(4) (2001) 979--986.
\newblock \href {http://arxiv.org/abs/https://doi.org/10.1021/ef010047l}
  {\path{arXiv:https://doi.org/10.1021/ef010047l}}, \href
  {https://doi.org/10.1021/ef010047l} {\path{doi:10.1021/ef010047l}}.
\newline\urlprefix\url{https://doi.org/10.1021/ef010047l}

\bibitem{li2017experimental}
X.~Li, Y.~Guo, E.~S. Boek, X.~Guo,
  \href{https://doi.org/10.1021/acs.energyfuels.7b01170}{Experimental study on
  kinetics of asphaltene aggregation in a microcapillary}, Energy \& Fuels
  31~(9) (2017) 9006--9015.
\newblock \href
  {http://arxiv.org/abs/https://doi.org/10.1021/acs.energyfuels.7b01170}
  {\path{arXiv:https://doi.org/10.1021/acs.energyfuels.7b01170}}, \href
  {https://doi.org/10.1021/acs.energyfuels.7b01170}
  {\path{doi:10.1021/acs.energyfuels.7b01170}}.
\newline\urlprefix\url{https://doi.org/10.1021/acs.energyfuels.7b01170}

\bibitem{ARCINIEGAS2014202}
L.~M. Arciniegas, T.~Babadagli,
  \href{https://www.sciencedirect.com/science/article/pii/S001623611400129X}{Asphaltene
  precipitation, flocculation and deposition during solvent injection at
  elevated temperatures for heavy oil recovery}, Fuel 124 (2014) 202--211.
\newblock \href {https://doi.org/https://doi.org/10.1016/j.fuel.2014.02.003}
  {\path{doi:https://doi.org/10.1016/j.fuel.2014.02.003}}.
\newline\urlprefix\url{https://www.sciencedirect.com/science/article/pii/S001623611400129X}

\bibitem{Hirschberg1984}
A.~Hirschberg, L.~deJong, B.~Schipper, J.~Meijer,
  \href{https://doi.org/10.2118/11202-PA}{{Influence of Temperature and
  Pressure on Asphaltene Flocculation}}, Society of Petroleum Engineers Journal
  24~(03) (1984) 283--293.
\newblock \href
  {http://arxiv.org/abs/https://onepetro.org/spejournal/article-pdf/24/03/283/2153769/spe-11202-pa.pdf}
  {\path{arXiv:https://onepetro.org/spejournal/article-pdf/24/03/283/2153769/spe-11202-pa.pdf}},
  \href {https://doi.org/10.2118/11202-PA} {\path{doi:10.2118/11202-PA}}.
\newline\urlprefix\url{https://doi.org/10.2118/11202-PA}

\bibitem{ZANGANEH2018633}
P.~Zanganeh, H.~Dashti, S.~Ayatollahi,
  \href{https://www.sciencedirect.com/science/article/pii/S001623611830005X}{Comparing
  the effects of ch4, co2, and n2 injection on asphaltene precipitation and
  deposition at reservoir condition: A visual and modeling study}, Fuel 217
  (2018) 633--641.
\newblock \href {https://doi.org/https://doi.org/10.1016/j.fuel.2018.01.005}
  {\path{doi:https://doi.org/10.1016/j.fuel.2018.01.005}}.
\newline\urlprefix\url{https://www.sciencedirect.com/science/article/pii/S001623611830005X}

\bibitem{ZANGANEH2015132}
P.~Zanganeh, H.~Dashti, S.~Ayatollahi,
  \href{https://www.sciencedirect.com/science/article/pii/S0016236115007541}{Visual
  investigation and modeling of asphaltene precipitation and deposition during
  co2 miscible injection into oil reservoirs}, Fuel 160 (2015) 132--139.
\newblock \href {https://doi.org/https://doi.org/10.1016/j.fuel.2015.07.063}
  {\path{doi:https://doi.org/10.1016/j.fuel.2015.07.063}}.
\newline\urlprefix\url{https://www.sciencedirect.com/science/article/pii/S0016236115007541}

\bibitem{Haji2014}
N.~Haji-Akbari, P.~Teeraphapkul, H.~S. Fogler,
  \href{https://doi.org/10.1021/ef4021125}{Effect of asphaltene concentration
  on the aggregation and precipitation tendency of asphaltenes}, Energy \&
  Fuels 28~(2) (2014) 909--919.
\newblock \href {http://arxiv.org/abs/https://doi.org/10.1021/ef4021125}
  {\path{arXiv:https://doi.org/10.1021/ef4021125}}, \href
  {https://doi.org/10.1021/ef4021125} {\path{doi:10.1021/ef4021125}}.
\newline\urlprefix\url{https://doi.org/10.1021/ef4021125}

\bibitem{VILASBOASFAVERO2017267}
C.~{Vilas Bôas Fávero}, T.~Maqbool, M.~Hoepfner, N.~Haji-Akbari, H.~S.
  Fogler,
  \href{https://www.sciencedirect.com/science/article/pii/S0001868616301920}{Revisiting
  the flocculation kinetics of destabilized asphaltenes}, Advances in Colloid
  and Interface Science 244 (2017) 267--280, special Issue in Honor of the 90th
  Birthday of Prof. Eli Ruckenstein.
\newblock \href {https://doi.org/https://doi.org/10.1016/j.cis.2016.06.013}
  {\path{doi:https://doi.org/10.1016/j.cis.2016.06.013}}.
\newline\urlprefix\url{https://www.sciencedirect.com/science/article/pii/S0001868616301920}

\bibitem{ENAYAT2020116250}
S.~Enayat, N.~{Rajan Babu}, J.~Kuang, S.~Rezaee, H.~Lu, M.~Tavakkoli, J.~Wang,
  F.~M. Vargas,
  \href{https://www.sciencedirect.com/science/article/pii/S0016236119316047}{On
  the development of experimental methods to determine the rates of asphaltene
  precipitation, aggregation, and deposition}, Fuel 260 (2020) 116250.
\newblock \href {https://doi.org/https://doi.org/10.1016/j.fuel.2019.116250}
  {\path{doi:https://doi.org/10.1016/j.fuel.2019.116250}}.
\newline\urlprefix\url{https://www.sciencedirect.com/science/article/pii/S0016236119316047}

\bibitem{AKBARZADEH2005159}
K.~Akbarzadeh, H.~Alboudwarej, W.~Y. Svrcek, H.~W. Yarranton,
  \href{https://www.sciencedirect.com/science/article/pii/S0378381205001159}{A
  generalized regular solution model for asphaltene precipitation from n-alkane
  diluted heavy oils and bitumens}, Fluid Phase Equilibria 232~(1) (2005)
  159--170.
\newblock \href {https://doi.org/https://doi.org/10.1016/j.fluid.2005.03.029}
  {\path{doi:https://doi.org/10.1016/j.fluid.2005.03.029}}.
\newline\urlprefix\url{https://www.sciencedirect.com/science/article/pii/S0378381205001159}

\bibitem{ROGEL2017271}
E.~Rogel, M.~Moir,
  \href{https://www.sciencedirect.com/science/article/pii/S0016236117308256}{Effect
  of precipitation time and solvent power on asphaltene characteristics}, Fuel
  208 (2017) 271--280.
\newblock \href {https://doi.org/https://doi.org/10.1016/j.fuel.2017.06.116}
  {\path{doi:https://doi.org/10.1016/j.fuel.2017.06.116}}.
\newline\urlprefix\url{https://www.sciencedirect.com/science/article/pii/S0016236117308256}

\bibitem{Maqbool2011}
T.~Maqbool, S.~Raha, M.~P. Hoepfner, H.~S. Fogler, Modeling the aggregation of
  asphaltene nanoaggregates in crude oil- precipitant systems, Energy \& Fuels
  25~(4) (2011) 1585--1596.

\bibitem{Ramos2020}
F.~Ramos-Pallares, D.~Santos, H.~W. Yarranton,
  \href{https://doi.org/10.1021/acs.energyfuels.0c01803}{Application of the
  modified regular solution model to crude oils characterized from a
  distillation assay}, Energy \& Fuels 34~(12) (2020) 15270--15284.
\newblock \href
  {http://arxiv.org/abs/https://doi.org/10.1021/acs.energyfuels.0c01803}
  {\path{arXiv:https://doi.org/10.1021/acs.energyfuels.0c01803}}, \href
  {https://doi.org/10.1021/acs.energyfuels.0c01803}
  {\path{doi:10.1021/acs.energyfuels.0c01803}}.
\newline\urlprefix\url{https://doi.org/10.1021/acs.energyfuels.0c01803}

\bibitem{duran2019}
J.~A. Duran, Y.~A. Casas, L.~Xiang, L.~Zhang, H.~Zeng, H.~W. Yarranton,
  \href{https://doi.org/10.1021/acs.energyfuels.8b03057}{Nature of asphaltene
  aggregates}, Energy \& Fuels 33~(5) (2019) 3694--3710.
\newblock \href
  {http://arxiv.org/abs/https://doi.org/10.1021/acs.energyfuels.8b03057}
  {\path{arXiv:https://doi.org/10.1021/acs.energyfuels.8b03057}}, \href
  {https://doi.org/10.1021/acs.energyfuels.8b03057}
  {\path{doi:10.1021/acs.energyfuels.8b03057}}.
\newline\urlprefix\url{https://doi.org/10.1021/acs.energyfuels.8b03057}

\bibitem{RODRIGUEZ2019116079}
S.~Rodriguez, E.~Baydak, F.~Schoeggl, S.~Taylor, G.~Hay, H.~Yarranton,
  \href{https://www.sciencedirect.com/science/article/pii/S0016236119314334}{Regular
  solution based approach to modeling asphaltene precipitation from native and
  reacted oils: Part 3, visbroken oils}, Fuel 257 (2019) 116079.
\newblock \href {https://doi.org/https://doi.org/10.1016/j.fuel.2019.116079}
  {\path{doi:https://doi.org/10.1016/j.fuel.2019.116079}}.
\newline\urlprefix\url{https://www.sciencedirect.com/science/article/pii/S0016236119314334}

\bibitem{gray2011}
M.~R. Gray, R.~R. Tykwinski, J.~M. Stryker, X.~Tan,
  \href{https://doi.org/10.1021/ef200654p}{Supramolecular assembly model for
  aggregation of petroleum asphaltenes}, Energy \& Fuels 25~(7) (2011)
  3125--3134.
\newblock \href {http://arxiv.org/abs/https://doi.org/10.1021/ef200654p}
  {\path{arXiv:https://doi.org/10.1021/ef200654p}}, \href
  {https://doi.org/10.1021/ef200654p} {\path{doi:10.1021/ef200654p}}.
\newline\urlprefix\url{https://doi.org/10.1021/ef200654p}

\bibitem{wang2010}
S.~Wang, J.~Liu, L.~Zhang, J.~Masliyah, Z.~Xu,
  \href{https://doi.org/10.1021/la9020004}{Interaction forces between
  asphaltene surfaces in organic solvents}, Langmuir 26~(1) (2010) 183--190,
  pMID: 19645456.
\newblock \href {http://arxiv.org/abs/https://doi.org/10.1021/la9020004}
  {\path{arXiv:https://doi.org/10.1021/la9020004}}, \href
  {https://doi.org/10.1021/la9020004} {\path{doi:10.1021/la9020004}}.
\newline\urlprefix\url{https://doi.org/10.1021/la9020004}

\bibitem{wang2009}
S.~Wang, J.~Liu, L.~Zhang, Z.~Xu, J.~Masliyah,
  \href{https://doi.org/10.1021/ef800812k}{Colloidal interactions between
  asphaltene surfaces in toluene}, Energy \& Fuels 23~(2) (2009) 862--869.
\newblock \href {http://arxiv.org/abs/https://doi.org/10.1021/ef800812k}
  {\path{arXiv:https://doi.org/10.1021/ef800812k}}, \href
  {https://doi.org/10.1021/ef800812k} {\path{doi:10.1021/ef800812k}}.
\newline\urlprefix\url{https://doi.org/10.1021/ef800812k}

\bibitem{maqbool2009}
T.~Maqbool, A.~T. Balgoa, H.~S. Fogler,
  \href{https://doi.org/10.1021/ef9002236}{Revisiting asphaltene precipitation
  from crude oils: A case of neglected kinetic effects}, Energy \& Fuels 23~(7)
  (2009) 3681--3686.
\newblock \href {http://arxiv.org/abs/https://doi.org/10.1021/ef9002236}
  {\path{arXiv:https://doi.org/10.1021/ef9002236}}, \href
  {https://doi.org/10.1021/ef9002236} {\path{doi:10.1021/ef9002236}}.
\newline\urlprefix\url{https://doi.org/10.1021/ef9002236}

\bibitem{Casas2019}
Y.~A. Casas, J.~A. Duran, F.~F. Schoeggl, H.~W. Yarranton,
  \href{https://doi.org/10.1021/acs.energyfuels.9b02571}{Settling of asphaltene
  aggregates in n-alkane diluted bitumen}, Energy \& Fuels 33~(11) (2019)
  10687--10703.
\newblock \href
  {http://arxiv.org/abs/https://doi.org/10.1021/acs.energyfuels.9b02571}
  {\path{arXiv:https://doi.org/10.1021/acs.energyfuels.9b02571}}, \href
  {https://doi.org/10.1021/acs.energyfuels.9b02571}
  {\path{doi:10.1021/acs.energyfuels.9b02571}}.
\newline\urlprefix\url{https://doi.org/10.1021/acs.energyfuels.9b02571}

\bibitem{balestrin2019}
L.~B. d.~S. Balestrin, R.~D. Francisco, C.~A. Bertran, M.~B. Cardoso, W.~Loh,
  \href{https://doi.org/10.1021/acs.energyfuels.9b00043}{Direct assessment of
  inhibitor and solvent effects on the deposition mechanism of asphaltenes in a
  brazilian crude oil}, Energy \& Fuels 33~(6) (2019) 4748--4757.
\newblock \href
  {http://arxiv.org/abs/https://doi.org/10.1021/acs.energyfuels.9b00043}
  {\path{arXiv:https://doi.org/10.1021/acs.energyfuels.9b00043}}, \href
  {https://doi.org/10.1021/acs.energyfuels.9b00043}
  {\path{doi:10.1021/acs.energyfuels.9b00043}}.
\newline\urlprefix\url{https://doi.org/10.1021/acs.energyfuels.9b00043}

\bibitem{HUFFMAN2021121320}
M.~Huffman, H.~S. Fogler,
  \href{https://www.sciencedirect.com/science/article/pii/S0016236121011996}{Asphaltene
  destabilization in the presence of dodecylbenzene sulfonic acid and
  dodecylphenol}, Fuel 304 (2021) 121320.
\newblock \href {https://doi.org/https://doi.org/10.1016/j.fuel.2021.121320}
  {\path{doi:https://doi.org/10.1016/j.fuel.2021.121320}}.
\newline\urlprefix\url{https://www.sciencedirect.com/science/article/pii/S0016236121011996}

\bibitem{meng2020}
J.~Meng, J.~B. You, X.~Zhang,
  \href{https://doi.org/10.1021/acs.jpcc.0c02220}{Viscosity-mediated growth and
  coalescence of surface nanodroplets}, The Journal of Physical Chemistry C
  124~(23) (2020) 12476--12484.
\newblock \href {http://arxiv.org/abs/https://doi.org/10.1021/acs.jpcc.0c02220}
  {\path{arXiv:https://doi.org/10.1021/acs.jpcc.0c02220}}, \href
  {https://doi.org/10.1021/acs.jpcc.0c02220}
  {\path{doi:10.1021/acs.jpcc.0c02220}}.
\newline\urlprefix\url{https://doi.org/10.1021/acs.jpcc.0c02220}

\bibitem{Zhang2015}
X.~Zhang, Z.~Lu, H.~Tan, L.~Bao, Y.~He, C.~Sun, D.~Lohse,
  \href{https://www.pnas.org/content/112/30/9253}{Formation of surface
  nanodroplets under controlled flow conditions}, Proceedings of the National
  Academy of Sciences 112~(30) (2015) 9253--9257.
\newblock \href
  {http://arxiv.org/abs/https://www.pnas.org/content/112/30/9253.full.pdf}
  {\path{arXiv:https://www.pnas.org/content/112/30/9253.full.pdf}}, \href
  {https://doi.org/10.1073/pnas.1506071112}
  {\path{doi:10.1073/pnas.1506071112}}.
\newline\urlprefix\url{https://www.pnas.org/content/112/30/9253}

\bibitem{sieben2015}
V.~J. Sieben, A.~K. Tharanivasan, J.~Ratulowski, F.~Mostowfi,
  \href{http://dx.doi.org/10.1039/C5LC00547G}{Asphaltenes yield curve
  measurements on a microfluidic platform}, Lab Chip 15 (2015) 4062--4074.
\newblock \href {https://doi.org/10.1039/C5LC00547G}
  {\path{doi:10.1039/C5LC00547G}}.
\newline\urlprefix\url{http://dx.doi.org/10.1039/C5LC00547G}

\bibitem{sieben2016}
V.~J. Sieben, A.~K. Tharanivasan, S.~I. Andersen, F.~Mostowfi,
  \href{https://doi.org/10.1021/acs.energyfuels.5b02216}{Microfluidic approach
  for evaluating the solubility of crude oil asphaltenes}, Energy \& Fuels
  30~(3) (2016) 1933--1946.
\newblock \href
  {http://arxiv.org/abs/https://doi.org/10.1021/acs.energyfuels.5b02216}
  {\path{arXiv:https://doi.org/10.1021/acs.energyfuels.5b02216}}, \href
  {https://doi.org/10.1021/acs.energyfuels.5b02216}
  {\path{doi:10.1021/acs.energyfuels.5b02216}}.
\newline\urlprefix\url{https://doi.org/10.1021/acs.energyfuels.5b02216}

\bibitem{mozaffari2021}
S.~Mozaffari, H.~Ghasemi, P.~Tchoukov, J.~Czarnecki, N.~Nazemifard,
  \href{https://doi.org/10.1021/acs.energyfuels.1c00717}{Lab-on-a-chip systems
  in asphaltene characterization: A review of recent advances}, Energy \& Fuels
  35~(11) (2021) 9080--9101.
\newblock \href
  {http://arxiv.org/abs/https://doi.org/10.1021/acs.energyfuels.1c00717}
  {\path{arXiv:https://doi.org/10.1021/acs.energyfuels.1c00717}}, \href
  {https://doi.org/10.1021/acs.energyfuels.1c00717}
  {\path{doi:10.1021/acs.energyfuels.1c00717}}.
\newline\urlprefix\url{https://doi.org/10.1021/acs.energyfuels.1c00717}

\bibitem{meng2021primary}
J.~Meng, J.~B. You, H.~Hao, X.~Tan, X.~Zhang,
  \href{https://www.sciencedirect.com/science/article/pii/S0016236121004609}{Primary
  submicron particles from early stage asphaltene precipitation revealed in
  situ by total internal reflection fluorescence microscopy in a model oil
  system}, Fuel 296 (2021) 120584.
\newblock \href {https://doi.org/https://doi.org/10.1016/j.fuel.2021.120584}
  {\path{doi:https://doi.org/10.1016/j.fuel.2021.120584}}.
\newline\urlprefix\url{https://www.sciencedirect.com/science/article/pii/S0016236121004609}

\bibitem{meng2021microfluidic}
J.~Meng, J.~B. You, G.~F. Arends, H.~Hao, X.~Tan, X.~Zhang, Microfluidic device
  coupled with total internal reflection microscopy for in situ observation of
  precipitation, The European Physical Journal E 44~(4) (2021) 1--8.
\newblock \href {https://doi.org/10.1140/epje/s10189-021-00066-1}
  {\path{doi:10.1140/epje/s10189-021-00066-1}}.

\bibitem{lu2017universal}
Z.~Lu, M.~H.~K. Schaarsberg, X.~Zhu, L.~Y. Yeo, D.~Lohse, X.~Zhang, Universal
  nanodroplet branches from confining the ouzo effect, Proceedings of the
  National Academy of Sciences 114~(39) (2017) 10332--10337.

\bibitem{Dyett2018Growth}
B.~Dyett, A.~Kiyama, M.~Rump, Y.~Tagawa, D.~Lohse, X.~Zhang,
  \href{http://dx.doi.org/10.1039/C8SM00705E}{Growth dynamics of surface
  nanodroplets during solvent exchange at varying flow rates}, Soft Matter 14
  (2018) 5197--5204.
\newblock \href {https://doi.org/10.1039/C8SM00705E}
  {\path{doi:10.1039/C8SM00705E}}.
\newline\urlprefix\url{http://dx.doi.org/10.1039/C8SM00705E}

\bibitem{Dyett2018Coalescence}
B.~Dyett, H.~Hao, D.~Lohse, X.~Zhang,
  \href{http://dx.doi.org/10.1039/C7SM02490H}{Coalescence driven
  self-organization of growing nanodroplets around a microcap}, Soft Matter 14
  (2018) 2628--2637.
\newblock \href {https://doi.org/10.1039/C7SM02490H}
  {\path{doi:10.1039/C7SM02490H}}.
\newline\urlprefix\url{http://dx.doi.org/10.1039/C7SM02490H}

\bibitem{Dyett2020}
B.~P. Dyett, X.~Zhang,
  \href{https://doi.org/10.1021/acsnano.0c03059}{Accelerated formation of h2
  nanobubbles from a surface nanodroplet reaction}, ACS Nano 14~(9) (2020)
  10944--10953, pMID: 32692921.
\newblock \href {http://arxiv.org/abs/https://doi.org/10.1021/acsnano.0c03059}
  {\path{arXiv:https://doi.org/10.1021/acsnano.0c03059}}, \href
  {https://doi.org/10.1021/acsnano.0c03059}
  {\path{doi:10.1021/acsnano.0c03059}}.
\newline\urlprefix\url{https://doi.org/10.1021/acsnano.0c03059}

\bibitem{Feng2013}
F.~Rao, Q.~Liu, \href{https://doi.org/10.1021/ef4016697}{Froth treatment in
  athabasca oil sands bitumen recovery process: A review}, Energy \& Fuels
  27~(12) (2013) 7199--7207.
\newblock \href {http://arxiv.org/abs/https://doi.org/10.1021/ef4016697}
  {\path{arXiv:https://doi.org/10.1021/ef4016697}}, \href
  {https://doi.org/10.1021/ef4016697} {\path{doi:10.1021/ef4016697}}.
\newline\urlprefix\url{https://doi.org/10.1021/ef4016697}

\bibitem{barton2017crc}
A.~F. Barton, \href{https://doi.org/10.1140/epje/s10189-021-00066-1}{CRC
  handbook of solubility parameters and other cohesion parameters}, Routledge,
  2017.
\newblock \href {https://doi.org/10.1201/9781315140575}
  {\path{doi:10.1201/9781315140575}}.
\newline\urlprefix\url{https://doi.org/10.1140/epje/s10189-021-00066-1}

\bibitem{ANGLE2006492}
C.~W. Angle, Y.~Long, H.~Hamza, L.~Lue,
  \href{https://www.sciencedirect.com/science/article/pii/S0016236105002644}{Precipitation
  of asphaltenes from solvent-diluted heavy oil and thermodynamic properties of
  solvent-diluted heavy oil solutions}, Fuel 85~(4) (2006) 492--506.
\newblock \href {https://doi.org/https://doi.org/10.1016/j.fuel.2005.08.009}
  {\path{doi:https://doi.org/10.1016/j.fuel.2005.08.009}}.
\newline\urlprefix\url{https://www.sciencedirect.com/science/article/pii/S0016236105002644}

\bibitem{wang2003asphaltene}
J.~Wang, J.~S. Buckley, \href{https://doi.org/10.1021/ef030030y}{Asphaltene
  stability in crude oil and aromatic solventsthe influence of oil
  composition}, Energy \& Fuels 17~(6) (2003) 1445--1451.
\newblock \href {http://arxiv.org/abs/https://doi.org/10.1021/ef030030y}
  {\path{arXiv:https://doi.org/10.1021/ef030030y}}, \href
  {https://doi.org/10.1021/ef030030y} {\path{doi:10.1021/ef030030y}}.
\newline\urlprefix\url{https://doi.org/10.1021/ef030030y}

\bibitem{BUBAKOVA2013540}
P.~Bubakova, M.~Pivokonsky, P.~Filip,
  \href{https://www.sciencedirect.com/science/article/pii/S0032591012007632}{Effect
  of shear rate on aggregate size and structure in the process of aggregation
  and at steady state}, Powder Technology 235 (2013) 540--549.
\newblock \href {https://doi.org/https://doi.org/10.1016/j.powtec.2012.11.014}
  {\path{doi:https://doi.org/10.1016/j.powtec.2012.11.014}}.
\newline\urlprefix\url{https://www.sciencedirect.com/science/article/pii/S0032591012007632}

\bibitem{WANG20181183}
Z.~Wang, J.~Nan, X.~Ji, Y.~Yang,
  \href{https://www.sciencedirect.com/science/article/pii/S0048969718310465}{Effect
  of the micro-flocculation stage on the flocculation/sedimentation process:
  The role of shear rate}, Science of The Total Environment 633 (2018)
  1183--1191.
\newblock \href
  {https://doi.org/https://doi.org/10.1016/j.scitotenv.2018.03.286}
  {\path{doi:https://doi.org/10.1016/j.scitotenv.2018.03.286}}.
\newline\urlprefix\url{https://www.sciencedirect.com/science/article/pii/S0048969718310465}

\bibitem{elimelech2013particle}
M.~Elimelech, J.~Gregory, X.~Jia, Particle deposition and aggregation:
  measurement, modelling and simulation, Butterworth-Heinemann, 2013.

\bibitem{subodhsen1999}
S.~Peramanu, P.~F. Clarke, B.~B. Pruden,
  \href{https://www.sciencedirect.com/science/article/pii/S0920410599000121}{Flow
  loop apparatus to study the effect of solvent, temperature and additives on
  asphaltene precipitation}, Journal of Petroleum Science and Engineering
  23~(2) (1999) 133--143.
\newblock \href {https://doi.org/https://doi.org/10.1016/S0920-4105(99)00012-1}
  {\path{doi:https://doi.org/10.1016/S0920-4105(99)00012-1}}.
\newline\urlprefix\url{https://www.sciencedirect.com/science/article/pii/S0920410599000121}

\bibitem{Mostowfi2009}
F.~Mostowfi, K.~Indo, O.~C. Mullins, R.~McFarlane,
  \href{https://doi.org/10.1021/ef8006273}{Asphaltene nanoaggregates studied by
  centrifugation}, Energy \& Fuels 23~(3) (2009) 1194--1200.
\newblock \href {http://arxiv.org/abs/https://doi.org/10.1021/ef8006273}
  {\path{arXiv:https://doi.org/10.1021/ef8006273}}, \href
  {https://doi.org/10.1021/ef8006273} {\path{doi:10.1021/ef8006273}}.
\newline\urlprefix\url{https://doi.org/10.1021/ef8006273}

\bibitem{Betancourt2009}
S.~S. Betancourt, G.~T. Ventura, A.~E. Pomerantz, O.~Viloria, F.~X. Dubost,
  J.~Zuo, G.~Monson, D.~Bustamante, J.~M. Purcell, R.~K. Nelson, R.~P. Rodgers,
  C.~M. Reddy, A.~G. Marshall, O.~C. Mullins,
  \href{https://doi.org/10.1021/ef800598a}{Nanoaggregates of asphaltenes in a
  reservoir crude oil and reservoir connectivity}, Energy \& Fuels 23~(3)
  (2009) 1178--1188.
\newblock \href {http://arxiv.org/abs/https://doi.org/10.1021/ef800598a}
  {\path{arXiv:https://doi.org/10.1021/ef800598a}}, \href
  {https://doi.org/10.1021/ef800598a} {\path{doi:10.1021/ef800598a}}.
\newline\urlprefix\url{https://doi.org/10.1021/ef800598a}

\bibitem{Rogel2011}
E.~Rogel, \href{https://doi.org/10.1021/ef100912b}{Effect of inhibitors on
  asphaltene aggregation: A theoretical framework}, Energy \& Fuels 25~(2)
  (2011) 472--481.
\newblock \href {http://arxiv.org/abs/https://doi.org/10.1021/ef100912b}
  {\path{arXiv:https://doi.org/10.1021/ef100912b}}, \href
  {https://doi.org/10.1021/ef100912b} {\path{doi:10.1021/ef100912b}}.
\newline\urlprefix\url{https://doi.org/10.1021/ef100912b}

\bibitem{Acevedo2005}
S.~Acevedo, L.~B. Gutierrez, G.~Negrin, J.~C. Pereira, B.~Mendez, F.~Delolme,
  G.~Dessalces, D.~Broseta, \href{https://doi.org/10.1021/ef040071+}{Molecular
  weight of petroleum asphaltenes: A comparison between mass spectrometry and
  vapor pressure osmometry}, Energy \& Fuels 19~(4) (2005) 1548--1560.
\newblock \href {http://arxiv.org/abs/https://doi.org/10.1021/ef040071+}
  {\path{arXiv:https://doi.org/10.1021/ef040071+}}, \href
  {https://doi.org/10.1021/ef040071+} {\path{doi:10.1021/ef040071+}}.
\newline\urlprefix\url{https://doi.org/10.1021/ef040071+}

\bibitem{Yarranton2000}
H.~W. Yarranton, H.~Alboudwarej, R.~Jakher,
  \href{https://doi.org/10.1021/ie000073r}{Investigation of asphaltene
  association with vapor pressure osmometry and interfacial tension
  measurements}, Industrial \& Engineering Chemistry Research 39~(8) (2000)
  2916--2924.
\newblock \href {http://arxiv.org/abs/https://doi.org/10.1021/ie000073r}
  {\path{arXiv:https://doi.org/10.1021/ie000073r}}, \href
  {https://doi.org/10.1021/ie000073r} {\path{doi:10.1021/ie000073r}}.
\newline\urlprefix\url{https://doi.org/10.1021/ie000073r}

\bibitem{Barrera2013}
D.~M. Barrera, D.~P. Ortiz, H.~W. Yarranton,
  \href{https://doi.org/10.1021/ef400142v}{Molecular weight and density
  distributions of asphaltenes from crude oils}, Energy \& Fuels 27~(5) (2013)
  2474--2487.
\newblock \href {http://arxiv.org/abs/https://doi.org/10.1021/ef400142v}
  {\path{arXiv:https://doi.org/10.1021/ef400142v}}, \href
  {https://doi.org/10.1021/ef400142v} {\path{doi:10.1021/ef400142v}}.
\newline\urlprefix\url{https://doi.org/10.1021/ef400142v}

\bibitem{YUDIN1998297}
I.~Yudin, G.~Nikolaenko, E.~Gorodetskii, V.~Kosov, V.~Melikyan, E.~Markhashov,
  D.~Frot, Y.~Briolant,
  \href{https://www.sciencedirect.com/science/article/pii/S0920410598000333}{Mechanisms
  of asphaltene aggregation in toluene–heptane mixtures}, Journal of
  Petroleum Science and Engineering 20~(3) (1998) 297--301.
\newblock \href {https://doi.org/https://doi.org/10.1016/S0920-4105(98)00033-3}
  {\path{doi:https://doi.org/10.1016/S0920-4105(98)00033-3}}.
\newline\urlprefix\url{https://www.sciencedirect.com/science/article/pii/S0920410598000333}

\bibitem{Haji2013}
N.~Haji-Akbari, P.~Masirisuk, M.~P. Hoepfner, H.~S. Fogler,
  \href{https://doi.org/10.1021/ef4001665}{A unified model for aggregation of
  asphaltenes}, Energy \& Fuels 27~(5) (2013) 2497--2505.
\newblock \href {http://arxiv.org/abs/https://doi.org/10.1021/ef4001665}
  {\path{arXiv:https://doi.org/10.1021/ef4001665}}, \href
  {https://doi.org/10.1021/ef4001665} {\path{doi:10.1021/ef4001665}}.
\newline\urlprefix\url{https://doi.org/10.1021/ef4001665}

\bibitem{ALMASI2021129624}
M.~Almasi,
  \href{https://www.sciencedirect.com/science/article/pii/S0022286020319384}{Molecular
  interactions and structural studies of toluene and (c5 - c10) 1-alkanol;
  mutual diffusion and virial coefficients}, Journal of Molecular Structure
  1230 (2021) 129624.
\newblock \href
  {https://doi.org/https://doi.org/10.1016/j.molstruc.2020.129624}
  {\path{doi:https://doi.org/10.1016/j.molstruc.2020.129624}}.
\newline\urlprefix\url{https://www.sciencedirect.com/science/article/pii/S0022286020319384}

\bibitem{zhu2016prediction}
Q.~Zhu, C.~D'Agostino, M.~Ainte, M.~Mantle, L.~Gladden, O.~Ortona, L.~Paduano,
  D.~Ciccarelli, G.~Moggridge, Prediction of mutual diffusion coefficients in
  binary liquid systems with one self-associating component from viscosity data
  and intra-diffusion coefficients at infinite dilution, Chemical Engineering
  Science 147 (2016) 118--127.

\bibitem{lemmonthermophysical}
E.~W. Lemmon, M.~O. McLinden, D.~G. Friend, Thermophysical properties of fluid
  systems, NIST chemistry webbook, Nat. Inst. Stand. Tech., Gaithersburg MD,
  20899, 2008.
\newblock \href {https://doi.org/https://doi.org/10.18434/T4D303}
  {\path{doi:https://doi.org/10.18434/T4D303}}.

\bibitem{gonccalves1987viscosity}
F.~Gon{\c{c}}alves, K.~Hamano, J.~Sengers, J.~Kestin, Viscosity of liquid
  toluene in the temperature range 25--75° c, International journal of
  thermophysics 8~(6) (1987) 641--647.

\end{thebibliography}

\begin{appendices}

\section{Appendix}
\renewcommand{\thefigure}{S\arabic{figure}}
\renewcommand{\thetable}{S\arabic{table}}

\subsection{Optical images of asphaltene particles precipitate from bitumen}

\begin{figure}[ht]
\centering
\includegraphics[width=1\columnwidth]{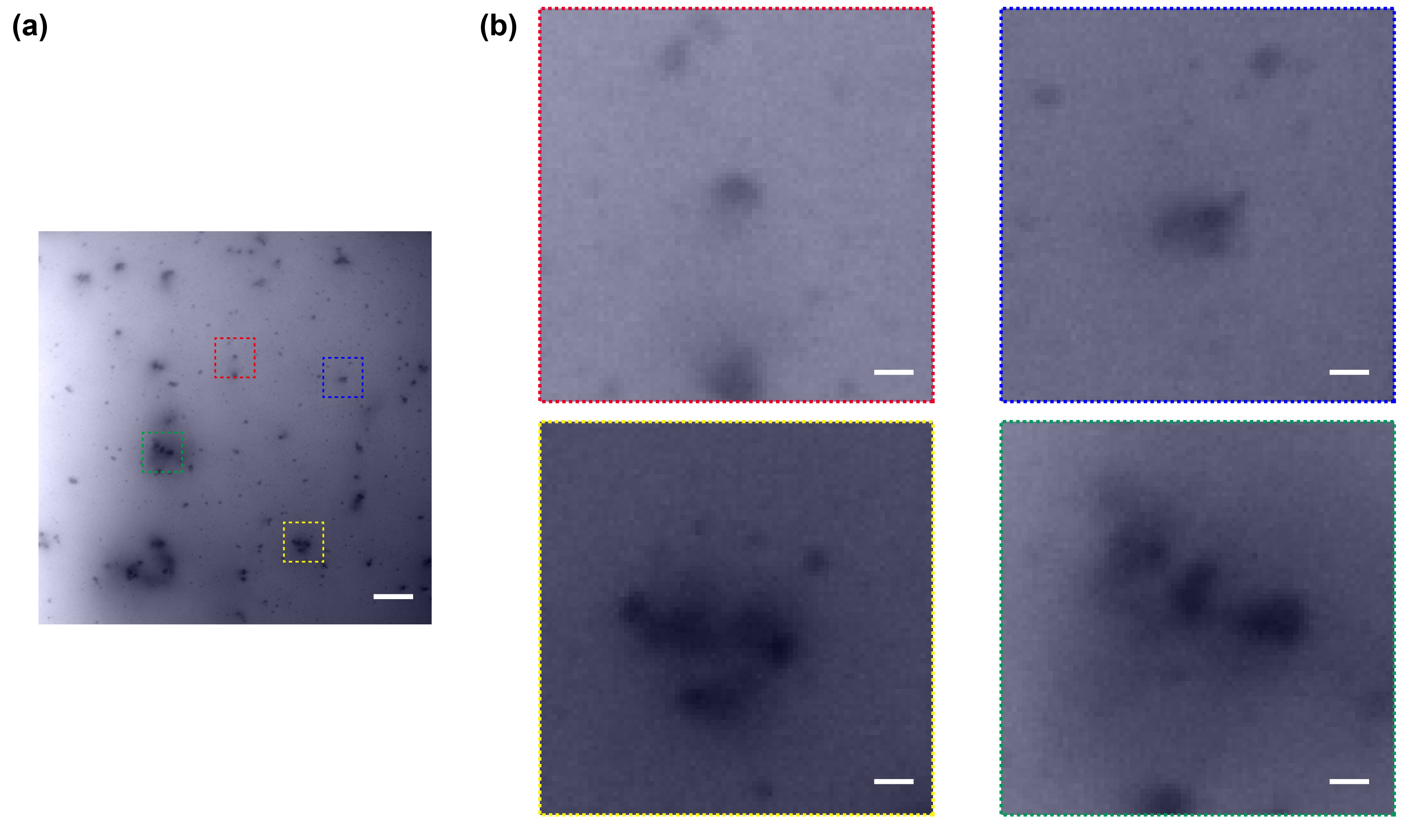}
\captionsetup{font={normal}}
\renewcommand{\captionfont}{\normalsize}
\caption{a) TIRF images of asphaltene particles at 5 $min$ from bitumen. Precipitation is induced by heptane. The length of the scale bar is 10 $\mu m$. b) Zoomed-in images of primary sub-micron particles in a) at locations with respective color boxes. The length of the scale bar is 1 $\mu m$. The images are false-colored. }
\label{bit_opt}
\end{figure}

\subsection{Chemical structure of the asphaltene particles precipitated in different types of solvents}

Confocal laser scanning microscope (Leica TCS SP5, Mannheim, Germany) was used to measure the emission spectrum of asphaltene particles. An Argon laser 
was used to excite the samples. The intensity of the fluorescence emission was measured at a 5 $nm$ interval from 500 $nm$ to 700 $nm$ and the fluorescence spectrum was plotted within the  range of wavelength.

\begin{figure}[ht]
\centering
\includegraphics[width=1\columnwidth]{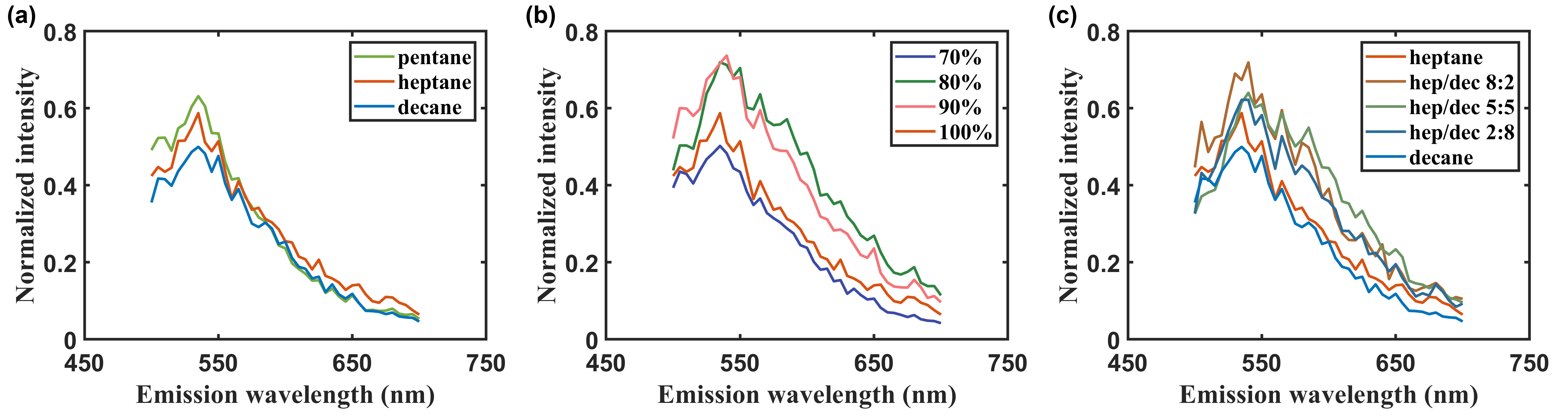}
\captionsetup{font={normal}}
\renewcommand{\captionfont}{\normalsize}
\caption{Fluorescence spectrum of asphaltene particles precipitated under a) different types of solvents, b)heptol with different heptane concentrations, and c) heptane-decane mixtures with different heptane concentrations.}
\label{spec}
\end{figure}

Figure \ref{spec} shows the fluorescence spectra for asphaltene particles precipitated in different types of alkanes and their mixture with toluene, as well as the mixture of heptane and decane. The spectra are normalized with three particles of each condition. Under 488 $nm$ excitation, all spectra show a wide band from 500 $nm$ to 700 $nm$. The wideband attributes to the chemical structure complexity of asphaltene molecules. 

The highest peak always appears at around 540 $nm$. Primary sub-micron particles and large aggregates have similar spectra, indicating the chemical structure of these two types of particles do not have different chemical structures.

\subsection{Estimation of mutual diffusion coefficients}

Mutual diffusivities of toluene-alkane binary mixtures were approximated using a modified Darken model \cite{ALMASI2021129624,zhu2016prediction}. 

\begin{figure}[ht]
\centering
\includegraphics[width=0.9\columnwidth]{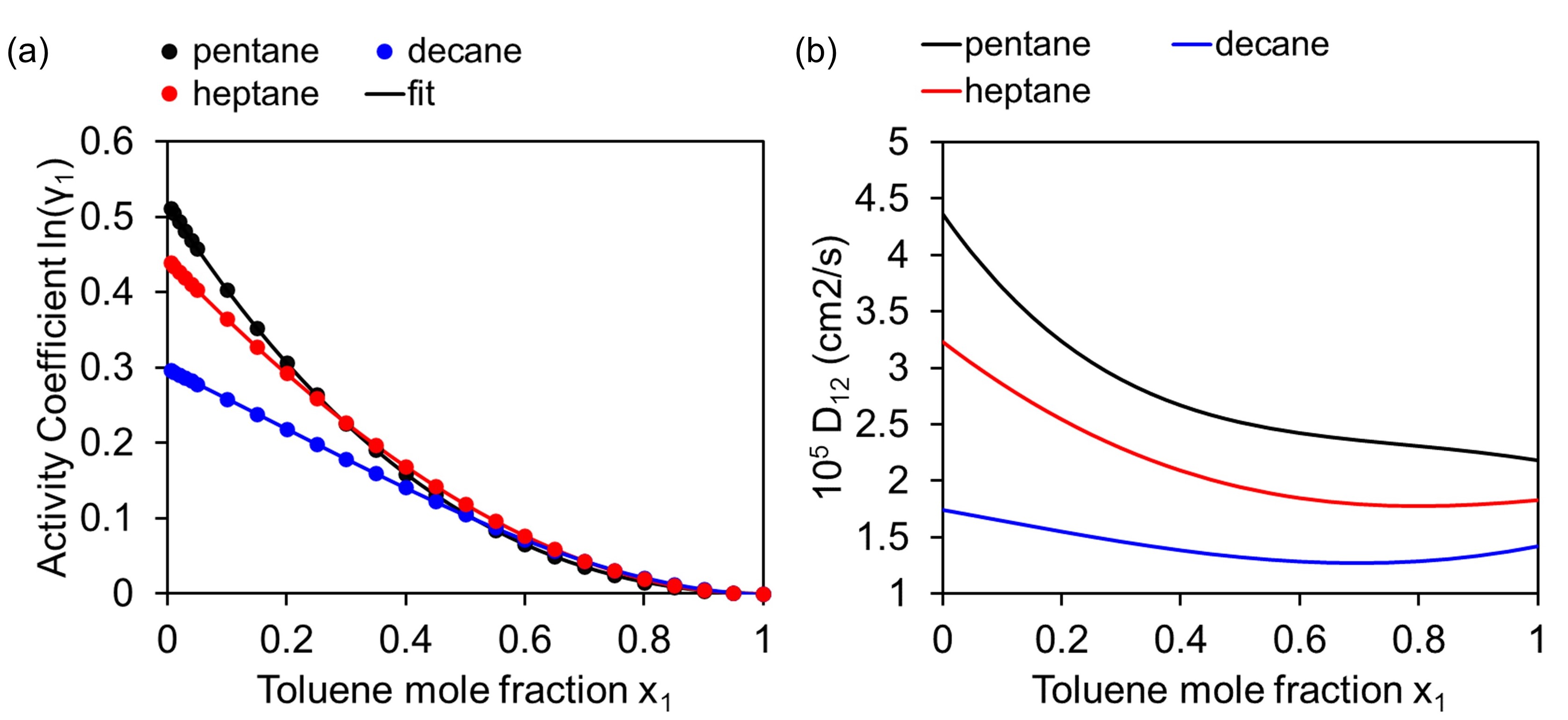}
\captionsetup{font={normal}}
\renewcommand{\captionfont}{\normalsize}
\caption{a) Calculated activity coefficients in binary toluene-alkane mixtures at 298.15 K with UNIFAC. b) Diffusion coefficients of toluene in different alkanes and mixture compositions.}
\label{diffusiongraph}
\end{figure}

\begin{equation}
    D_{ij} = \left(x_jD_{i}^{*} + x_iD_{j}^{*} \right) \Gamma
\label{equationDarken}
\end{equation}

The Darken model includes a thermodynamic correction factor ($\Gamma$) which considers the change in activity coefficient ($\gamma_i$) of the solute within a chemical potential gradient. The activity coefficient values were obtained at $T = 298.15 K$ using the UNIFAC thermodynamic model within the software package Symmetry (Figure \ref{diffusiongraph}a). Values were fit to polynomial equations to obtain their derivative for $\Gamma$ calculations as a function of mixture composition.

\begin{equation}
    \Gamma = 1 + x_i \left( \frac{\partial \ln \gamma_i}{\partial x_i} \right)
\label{equationthermo}
\end{equation}

The self-diffusion coefficients ($D_{1}^{*}$ and $D_{2}^{*}$) in the Darken model (Equation \ref{equationDarken}) were approximated using the Wilke-Chang equation for binary liquids (Equation \ref{Wilke-Chang}). $M_j$, $T$, $\mu_{j}$ and $V_{i,BP}$ represent the molecular weight of the solvent, temperature, solvent viscosity, and molar volume of solute at normal boiling point conditions, respectively. The pure compound properties summarized in Table \ref{Pure compound table} were obtained from the NIST database. The association factor $\psi_j = 1.0$ for unassociated solvents.

\begin{equation}
    D_{i}^{*} = \frac{7.4 \times 10^{-8} \left(\psi_jM_j \right)^{0.5}T}{\mu_{j} \times V_{i,BP}^{0.6}}
\label{Wilke-Chang}
\end{equation}

\begin{table}[ht]
\captionsetup{font = {small}}
\caption{Pure compound properties obtained from NIST for each species \cite{lemmonthermophysical,gonccalves1987viscosity}.}
\centering
\begin{tabular}{|c| c| c| c| c| c|}
\hline
Compound & Formula & M & $T_{BP}$ & $V_{BP}$ & $\mu$\\ & & (g/mol) & $(K)$ & (${cm^3}/mol$) & $(cp)$ \\
\hline
Toluene & $C_{7}H_{8}$ & 92.14 & 383.8 & 118.26 & 0.558 \\
\hline
Pentane & $C_{5}H_{12}$ & 72.15 & 309.2 & 118.33 & 0.245\\
\hline
Heptane & $C_{7}H_{16}$ & 100.20 & 371.5 & 163.14 & 0.390\\
\hline
Decane & $C_{10}H_{22}$ & 142.28 & 447.3 & 235.60 & 0.848\\
\hline
\end{tabular}
\label{Pure compound table}
\end{table}

\subsection*{Nomenclature}
$\gamma_i$ = Activity coefficient of solute at T and $x_i$

$x_i$ = Molar composition of solute $i$

$x_j$ = Molar composition of solvent $j$

$D_{i}^{*}$ = Self-diffusion coefficient of solute $i$ ($cm^2 \cdot s^{-1}$)

$D_{j}^{*}$ = Self-diffusion coefficient of solvent $j$ ($cm^2 \cdot s^{-1}$)

$D_{ij}$ = Mutual diffusion coefficient of solute $i$ in solvent $j$ ($cm^2 \cdot s^{-1}$)

$\Gamma$ = Thermodynamic factor

$M_j$ = Molecular weight of solvent $j$ ($g/mol$)

$T$ = Temperature ($K$)

$\mu_j$ = Viscosity of solvent $j$ at T ($cp$)

$\psi_j$ = Association factor of solvent $j$

$V_{i,BP}$ = Molar volume of solute $i$ at normal boiling temperature ($cm^3 \cdot mol^{-1}$)

\end{appendices}
\end{document}